\documentclass[fleqn,usenatbib]{mnras}

\usepackage{newtxtext,newtxmath}

\usepackage[T1]{fontenc}
\usepackage{ae,aecompl}

\usepackage{graphicx}	
\usepackage{amsmath}	
\usepackage{amssymb}	
\usepackage{siunitx}

\title[Machine Learning for Atmospheric Retrieval]{Assessment of Supervised Machine Learning for Atmospheric Retrieval of Exoplanets}

\author[M. C. Nixon and N. Madhusudhan]{
Matthew C. Nixon\thanks{E-mail: mnixon@ast.cam.ac.uk}
and Nikku Madhusudhan\thanks{E-mail: nmadhu@ast.cam.ac.uk}
\\
Institute of Astronomy, University of Cambridge, Madingley Road, Cambridge CB3 0HA, UK
}

\date{Accepted XXX. Received YYY; in original form ZZZ}

\pubyear{2020}

\begin{document}
\label{firstpage}
\pagerange{\pageref{firstpage}--\pageref{lastpage}}
\maketitle

\begin{abstract}
Atmospheric retrieval of exoplanets from spectroscopic observations requires an extensive exploration of a highly degenerate and high-dimensional parameter space to accurately constrain atmospheric parameters. Retrieval methods commonly conduct Bayesian parameter estimation and statistical inference using sampling algorithms such as Markov Chain Monte Carlo (MCMC) or Nested Sampling. Recently several attempts have been made to use machine learning algorithms either to complement or replace fully Bayesian methods. While much progress has been made, these approaches are still at times unable to accurately reproduce results from contemporary Bayesian retrievals. The goal of our present work is to investigate the efficacy of machine learning for atmospheric retrieval. As a case study, we use the Random Forest supervised machine learning algorithm which has been applied previously with some success for atmospheric retrieval of the hot Jupiter WASP-12b using its near-infrared transmission spectrum. We reproduce previous results using the same approach and the same semi-analytic models, and subsequently extend this method to develop a new algorithm that results in a closer match to a fully Bayesian retrieval. We combine this new method with a fully numerical atmospheric model and demonstrate excellent agreement with a Bayesian retrieval of the transmission spectrum of another hot Jupiter, HD~209458b. Despite this success, and achieving high computational efficiency, we still find that the machine learning approach is computationally prohibitive for high-dimensional parameter spaces that are routinely explored with Bayesian retrievals with modest computational resources. We discuss the trade offs and potential avenues for the future.  
\end{abstract}

\begin{keywords}
methods: data analysis -- planets and satellites: atmospheres
\end{keywords}

\defcitealias{PMN18}{MN18}



\section{Introduction}

Machine learning and artificial intelligence are becoming increasingly prevalent in many areas of astrophysics. Many popular machine learning techniques have been applied to astrophysical problems including galaxy classification \citep{Banerji_10}, characterisation of supernovae \citep{lochner_16}, and exoplanet detection \citep{Shallue_18}. Recently a number of attempts have been made to use machine learning to retrieve properties of exoplanet atmospheres from spectroscopic data. \citet{waldmann_2016} trained a Deep Belief Neural Network to make qualitative predictions about which molecular and atomic opacity sources to include in a traditional retrieval framework. \citet{PMN18} employed a supervised learning algorithm called Random Forest to retrieve atmospheric properties of the hot giant planet WASP-12b. \citet{zingales_2018} developed a Generative Adversarial Network which uses unsupervised learning to predict planetary parameters as well as atomic and molecular abundances. \citet{Soboczenski_2018} explored the use of Deep Neural Networks to make inference from synthetic spectra of terrestrial planets and incorporated Monte Carlo dropout in order to approximate model uncertainty. This method was further developed in \citet{Cobb_19}, who used an ensemble of Neural Networks and incorporated domain-specific knowledge to improve performance. 

A limitation of applying machine learning for retrievals has been the statistical interpretation of parameter predictions given the observed data. Traditionally, atmospheric retrieval has used Bayesian inference techniques to estimate the central values and uncertainties of the model parameters which fit an observed spectrum \citep{Madhu_18}. Such techniques used for both transmission and emission spectra include MCMC \citep[e.g.,][]{MS10,Line_13,Cubillos_13} and Nested Sampling \citep[e.g.,][]{Benneke_13,Waldmann_15,Oreshenko_17,Sid_18}. When applied to atmospheric spectra, retrievals have often highlighted strong degeneracies between model parameters \citep[e.g.,][]{Benneke_12,Griffith_14,Line_16,Welbanks_19}. It is therefore important when carrying out a retrieval to use a method that is able to find these model degeneracies and accurately capture the inherent uncertainties in the observed spectra. Previous studies employing machine learning have produced either a set of predictions similar to running an ensemble of optimal estimation procedures \citep{zingales_2018} or an approximation of the posterior distribution that is not shown to match the result of a Bayesian inference procedure \citep{Cobb_19}. In cases where attempts were made to compare a machine learning retrieval with a Bayesian retrieval \citep{zingales_2018,PMN18}, the posterior distributions between the retrievals reveal some discrepancies, as discussed later in this work.

In this paper we focus on supervised ensemble learning, similar to that employed by \citet{PMN18}, referred to as \citetalias{PMN18} hereafter. \citetalias{PMN18} use the Random Forest algorithm, to train multiple estimators (or trees) to predict parameter values which best describe the transmission spectrum of WASP-12b. The distribution of predictions made by the estimators is used to find the uncertainties on the estimated parameter values. The results of their Random Forest retrieval are compared to a Nested Sampling retrieval, and whilst the two retrievals yield comparable parameter estimates, the uncertainties are not consistent between the two methods. Our goal in this study is to determine if it is possible to develop a more statistically sound retrieval framework using the Random Forest algorithm.

In Section \ref{section:methods} we first reproduce the retrieval results of \citetalias{PMN18} (using both Nested Sampling and Random Forest), using the WASP-12b transmission spectrum. To this end, we use the same semi-analytic model used in \citetalias{PMN18}. We then modify and extend the Random Forest method to perform a retrieval of the same spectrum that produces results whose uncertainties are closer to those found in a Nested Sampling retrieval. In Section \ref{section:HD} we combine this extended Random Forest method with the fully numerical forward model described in \citet{pinhas_18}. We validate our algorithm using synthetic spectra before conducting a case study of the Hubble Wide-Field Camera 3 (WFC3) transmission spectrum of HD~209458b, once again comparing the results of Random Forest and Nested Sampling retrievals. In Section \ref{section:discussion} we discuss the difficulties of applying these methods to more complex cases which would require a larger parameter space to be explored than previous machine learning retrievals. We also examine more generally the possible benefits and drawbacks of incorporating machine learning, particularly ensemble learning as explored in this paper, into the retrieval process.


\section{Methods} \label{section:methods}

\begin{figure}
\includegraphics[width=\columnwidth]{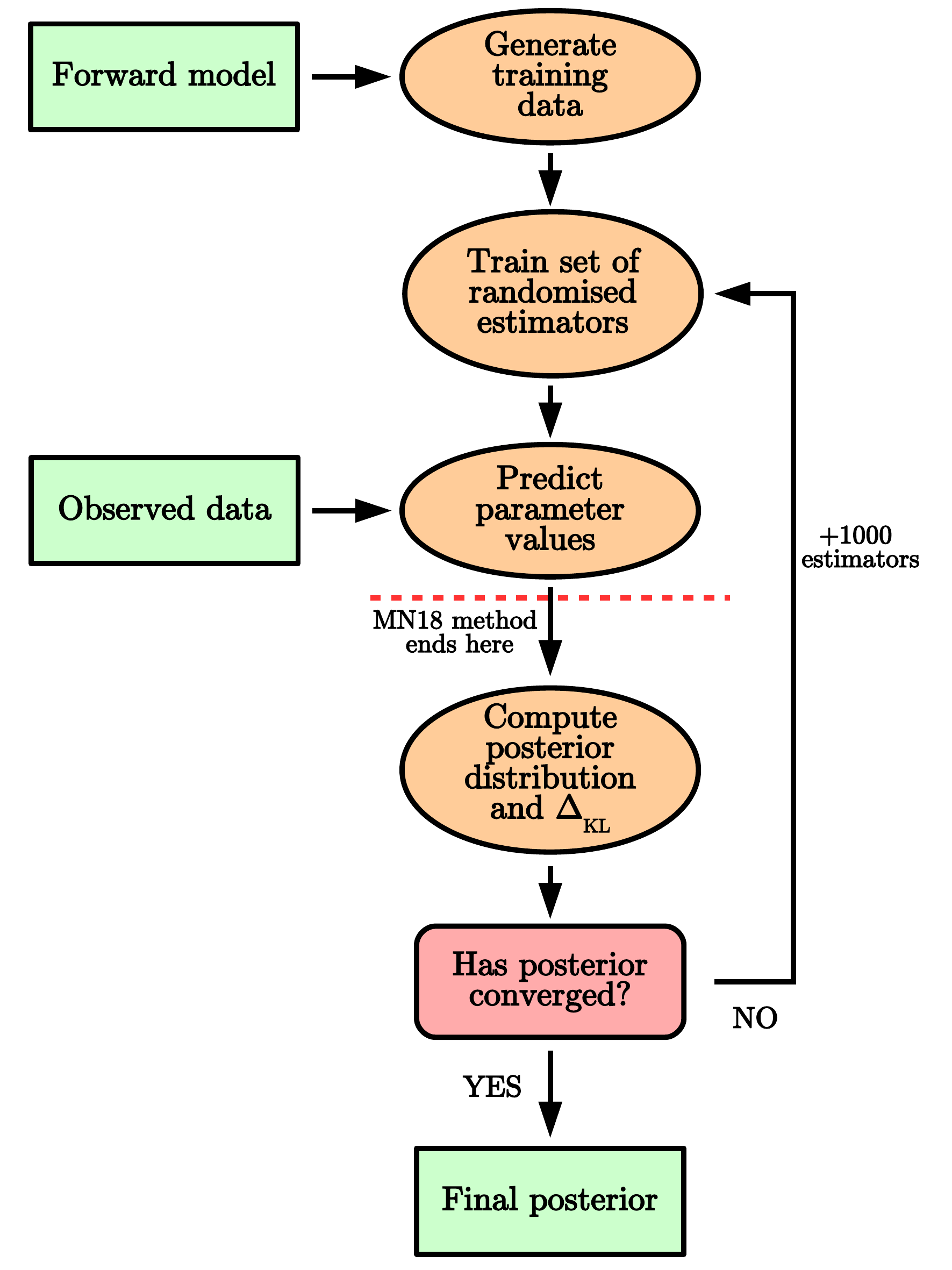}
    \caption{Flowchart describing the extended Random Forest retrieval framework. As indicated, the key differences between this and the \citetalias{PMN18} method are the calculation of the posterior distribution using the predicted parameter values and the iterative process of adding more trees until the posterior converges.}
    \label{fig:rf_flowchart}
\end{figure}

\subsection{Reproduction of previous results}

We begin by reproducing the results of \citetalias{PMN18}. We consider the same observed data as that paper, namely the WFC3 transmission spectrum of WASP-12b \citep{kreidberg_2015}. This spectrum consists of 13 binned data points in the infrared, at wavelengths ranging from $0.84-1.67$ $\mu$m. In order to retrieve atmospheric properties from the spectrum, two components are required: a forward model to calculate a transmission spectrum from a given set of parameters describing the atmospheric structure and composition, and a parameter estimation algorithm which finds the values of the model parameters that best fit the observed data. In this section we adopt the forward model of \citet{HK17} for consistency with the previous study. This semi-analytic model is used to produce a binned spectrum at the wavelengths of the WFC3 data given the values of five parameters: isotherm temperature $T_{\textnormal{iso}}$, the abundances of H$_2$O, HCN and NH$_3$, and a parameter to describe cloud opacity, $\kappa_0$. For the parameter estimation we follow the two approaches considered in \citetalias{PMN18}, first using the Nested Sampling algorithm MultiNest \citep{multinest2}, specifically its Python implementation PyMultiNest \citep{Buchner_2014}, and then using the implementation of the Random Forest algorithm \citep{random_forest} from scikit-learn.

\subsubsection{Nested Sampling}

In a traditional retrieval, Bayesian inference is used to estimate the values of model parameters given some observed data. Suppose we want to find the probability distribution of a set of parameters, denoted $\theta$, given some observed data, $d$. We can express this using Bayes' theorem:

\begin{equation}
    p(\theta|d) = \frac{p(d|\theta)p(\theta)}{p(d)}.
\end{equation}
Typically $p(\theta|d)$ is called the posterior, $p(d|\theta)$ is called the likelihood and is denoted $\mathcal{L}$, $p(\theta)$ is called the prior, and $p(d)$ is called the Bayesian evidence and is denoted $\mathcal{Z}$. Since $\mathcal{Z}$ does not depend on $\theta$, it simply acts as a normalisation factor and therefore is not needed for parameter estimation, however it can be used to compare different models.

Nested Sampling \citep{skilling_04} is a Monte Carlo algorithm designed to efficiently compute the Bayesian evidence of a model. It is also highly effective at sampling complex multimodal posterior distributions and is commonly used in many retrieval frameworks \citep[e.g.,][]{Benneke_13,Sid_18}. The algorithm initially selects a number of live points drawn from the defined prior volume, and evaluates the likelihoods of these points. Assuming Gaussian uncertainty on the measurements of the spectral data points, the likelihood is defined as

\begin{equation}
    \mathcal{L} = \mathcal{L}_0 \exp \bigg( - \frac{\chi^2}{2} \bigg),
    \label{eq:like}
\end{equation}
with
\begin{equation}
    \chi^2 = \sum_i \frac{(\hat{y}_i-\bar{y}_i)^2}{\sigma_i^2},
    \label{eq:chi}
\end{equation}
where $\bar{y}_i$ and $\sigma_i$ are the mean and standard deviation of the observed data point $i$, and $\hat{y}_i$ is the model prediction for data point $i$.

Having calculated $\mathcal{L}$ for each live point, the point with the lowest likelihood is discarded and replaced by a new one with a higher likelihood. This means that the volume contained within the set of live points continually shrinks, with the minimum likelihood bound by the volume progressively increasing. This process continues and $\mathcal{Z}$ is calculated until converging to within some pre-defined tolerance. Since the evidence calculation requires a thorough sampling of the parameter space, the Nested Sampling algorithm can therefore be used to estimate posterior distributions.

Using MultiNest in conjunction with the forward model described in \citet{HK17}, we reproduce the results from the Nested Sampling retrieval shown in \citetalias{PMN18}. The retrieved values and posterior distributions from this retrieval are shown in Figure \ref{fig:pmn_ns}. We obtain some constraints on the H$_2$O abundance while the HCN and NH$_3$ abundances remain unconstrained. The value of $\kappa_0$ is constrained to within 2 dex. The retrieved parameter values and associated uncertainties are consistent with the \citetalias{PMN18} Nested Sampling retrieval (see table 1 of that paper).

\begin{table}
	\centering
	\caption{Description of priors for retrievals of the transmission spectrum of WASP-12b. The priors have the same form for all chemical abundances $X_i$.}
	\hfill \\
	\label{tab:priors_wasp}
	\def\arraystretch{1.5}
	\setlength{\arrayrulewidth}{1.3pt}
	\begin{tabular}{cccc}
		\hline
		Parameter & Lower Bound & Upper Bound & Prior \\
		\hline
		$T_{\textnormal{iso}}$ (K) & $500$ & $2900$ & uniform \\
		$X_i$ & $10^{-13}$ & $1$ & log-uniform\\
		$\kappa_0$ (cm$^2$ g$^{-1}$) & $10^{-13}$ & $10^2$ & log-uniform\\
		\hline
	\end{tabular}
\end{table}

\begin{figure*}
\includegraphics[width=0.7\textwidth]{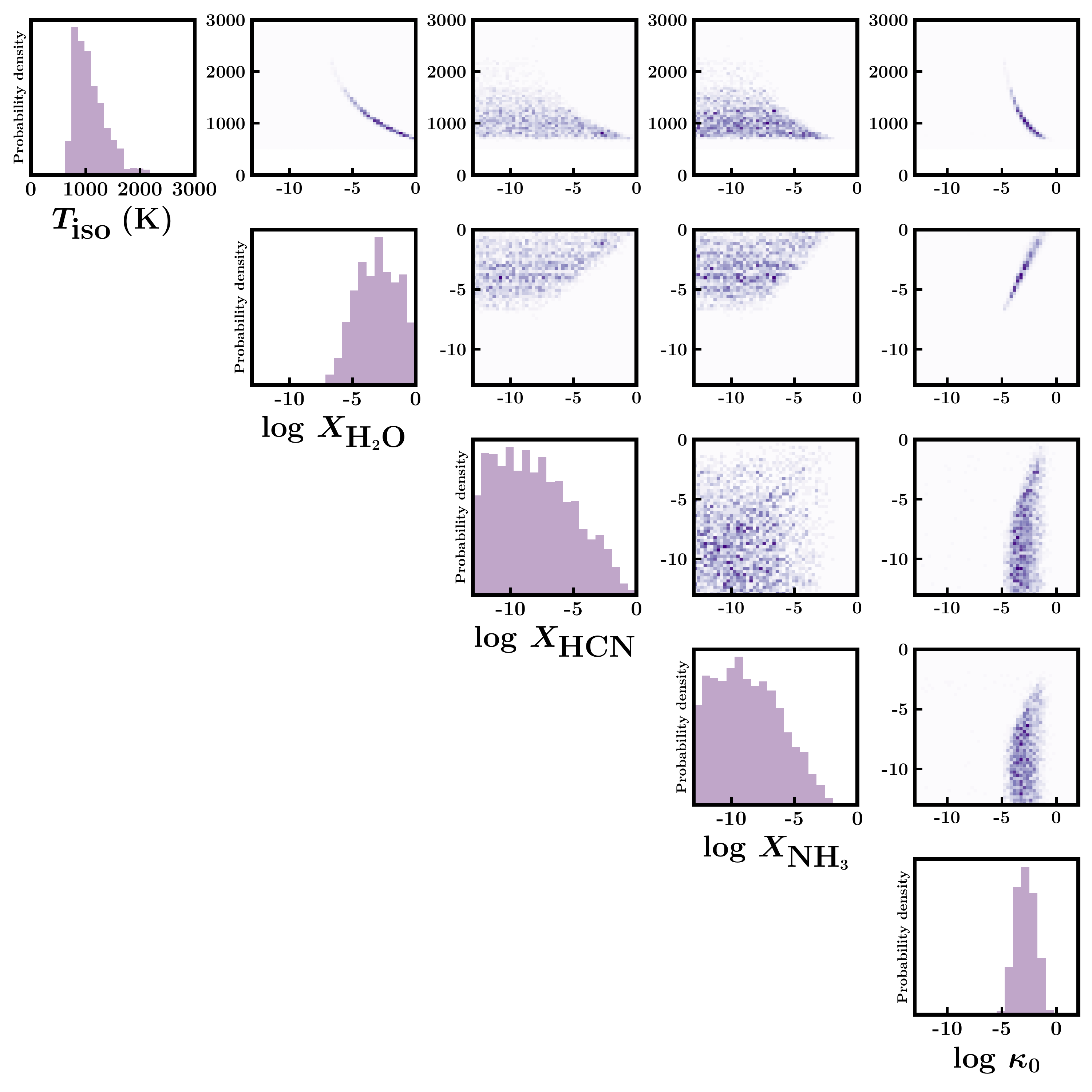}

	\def\arraystretch{1.5}
	\setlength{\arrayrulewidth}{1.3pt}
	\centering
	\Large{
	\vspace{-5.2cm}\hspace{-6.8cm}\begin{tabular}{cc}
		\hline
		Parameter & Value $^{+1\sigma}_{-1\sigma}$ \\
		\hline
		$T_{\textnormal{iso}}$ (K) & $1030^{+367}_{-224}$ \\		
		$\log X_{\textnormal{H}_2\textnormal{O}}$ & $-3.04^{+1.86}_{-1.84}$ \\
		$\log X_{\textnormal{HCN}}$ & $-8.36^{+3.66}_{-3.10}$ \\
		$\log X_{\textnormal{NH}_3}$ & $-8.97^{+3.01}_{-2.59}$ \\
		$\log \kappa_0$ & $-2.86^{+0.98}_{-0.96}$ \\
		\hline
	\end{tabular}}
	\vspace{0.5cm}
	
    \caption{Posterior distributions of Nested Sampling retrieval of the WFC3 transmission spectrum of WASP-12b following the methods of \citetalias{PMN18}. Inset: retrieved parameter values and associated 1$\sigma$ uncertainties.}
    \label{fig:pmn_ns}
\end{figure*}

\subsubsection{Random Forest}

Random Forest is a supervised machine learning algorithm. Supervised algorithms are trained on a data set that has been labelled in some way, and then try to predict the labels of some new, unlabelled data. The Random Forest method stems from the older decision tree algorithm \citep{decision_trees}. Decision trees, and hence Random Forest, can be used for either classification or regression tasks; here we outline its application to a general regression problem, since this is how the algorithm is applied to atmospheric retrieval.

We define the feature space $\mathcal{X}$ to be the vector space containing all possible input samples (binned spectra). The dimension of $\mathcal{X}$ is equal to the number of features in a sample, i.e. the number of data points in a single spectrum $\textbf{x}$. Similarly, we can define the space of all possible output labels $\textbf{y}$ (free parameters in the forward model) as $\mathcal{Y}$. In this context, the supervised machine learning problem becomes equivalent to finding the best possible partition of $\mathcal{X}$, where each partition corresponds to a different set of parameter estimates. The decision tree algorithm works by partitioning $\mathcal{X}$ into subspaces and assigning different values from $\mathcal{Y}$ to each subspace. In order to describe this further, we introduce some definitions from graph theory:

\begin{itemize}
    \item A graph is a collection of nodes and edges, where a node can be connected to another node by an edge.
    \item A graph can be either undirected, meaning that if there is an edge from node $a$ to node $b$ then there is automatically an edge from node $b$ to node $a$, or directed if this is not the case.
    \item If there is an edge from node $a$ to node $b$ but not from $b$ to $a$, then $a$ is said to be the parent of $b$ and $b$ the child of $a$.
    \item A tree is a graph in which there is exactly one path between any two nodes.
    \item If there exists a node in a tree where all edges are directed away from that node, then the node is called the root.
    \item A node in a tree which has no child nodes is called a leaf.
\end{itemize}
A decision tree can be defined as a directed tree in which any node $n$ corresponds to some subspace $\mathcal{X}_n$ of the feature space, with a root node that represents the entire space $\mathcal{X}$. Each leaf in the tree is assigned a value from the output space $\mathcal{Y}$. The aim of the learning process is therefore to determine the tree structure which best captures the relationship between the spaces $\mathcal{X}$ and $\mathcal{Y}$. We can quantify the concept of how well the model captures the relationship by defining the impurity $I_n$ of a node $n$:

\begin{equation}
    I_n = \frac{1}{N_n} \sum_{\textbf{x} \in \mathcal{X}_n} | \textbf{y}(\textbf{x}) - \hat{\textbf{y}}_n(\textbf{x}) |^2,
\end{equation}
where $N_n$ is the number of samples in the training data set which are found in the subspace $\mathcal{X}_n$, $\textbf{y}(\textbf{x})$ is the true value of the label corresponding to the sample $\textbf{x}$, and $\hat{\textbf{y}}_n(\textbf{x})$ is the value of the label for $\textbf{x}$ currently predicted by the model. The impurity is similar to the $\chi^2$ metric of equation \ref{eq:chi}. The algorithm proceeds by considering existing leaf nodes in the tree and splitting them into two or more child nodes (thus further partitioning the data set) such that the decrease in impurity from the parent node to the child nodes is maximised. This continues until some pre-determined tolerance in the impurity decrease is reached. 

The Random Forest algorithm is an ensemble method which uses a large set of decision trees. Ensemble methods aim to improve the robustness of predictions by training multiple models that have been randomly perturbed in some way. The ensemble prediction is then a combination of the individual model predictions. Randomness can be introduced in two ways: by training each tree on a random subset of the full training data set, which is sampled with replacement, and by limiting each tree to train using a random subset of features. It can be shown that an ensemble of randomised decision trees produces a more robust prediction than using a single tree \citep[see for example Chapter 4 of][]{louppe_thesis}.

In order to reproduce the Random Forest results of \citetalias{PMN18}, we use the forward model of \citet{HK17} to generate a training set of 100000 noisy synthetic WFC3 spectra in the wavelength range $0.8-1.7$ $\mu$m. For each spectrum, the value of each free parameter is chosen at random from a uniform or log-uniform distribution from within the prior ranges specified in \citetalias{PMN18} (see Table \ref{tab:priors_wasp}). The planetary and stellar radii are fixed at $R_{p} = 1.79R_J$ and $R_* = 1.577R_{\odot}$. We produce the training set at a higher wavelength resolution and larger wavelength range than the WFC3 spectrum, opting for $R=2000$ between 0.2 and 2.0 $\mu$m. This approach allows us to use the same training data set for multiple observation instances of the planet and would reduce the overall computation time of our method if other spectra of the same planet were to be analysed.

\begin{figure*}
\includegraphics[width=0.7\textwidth]{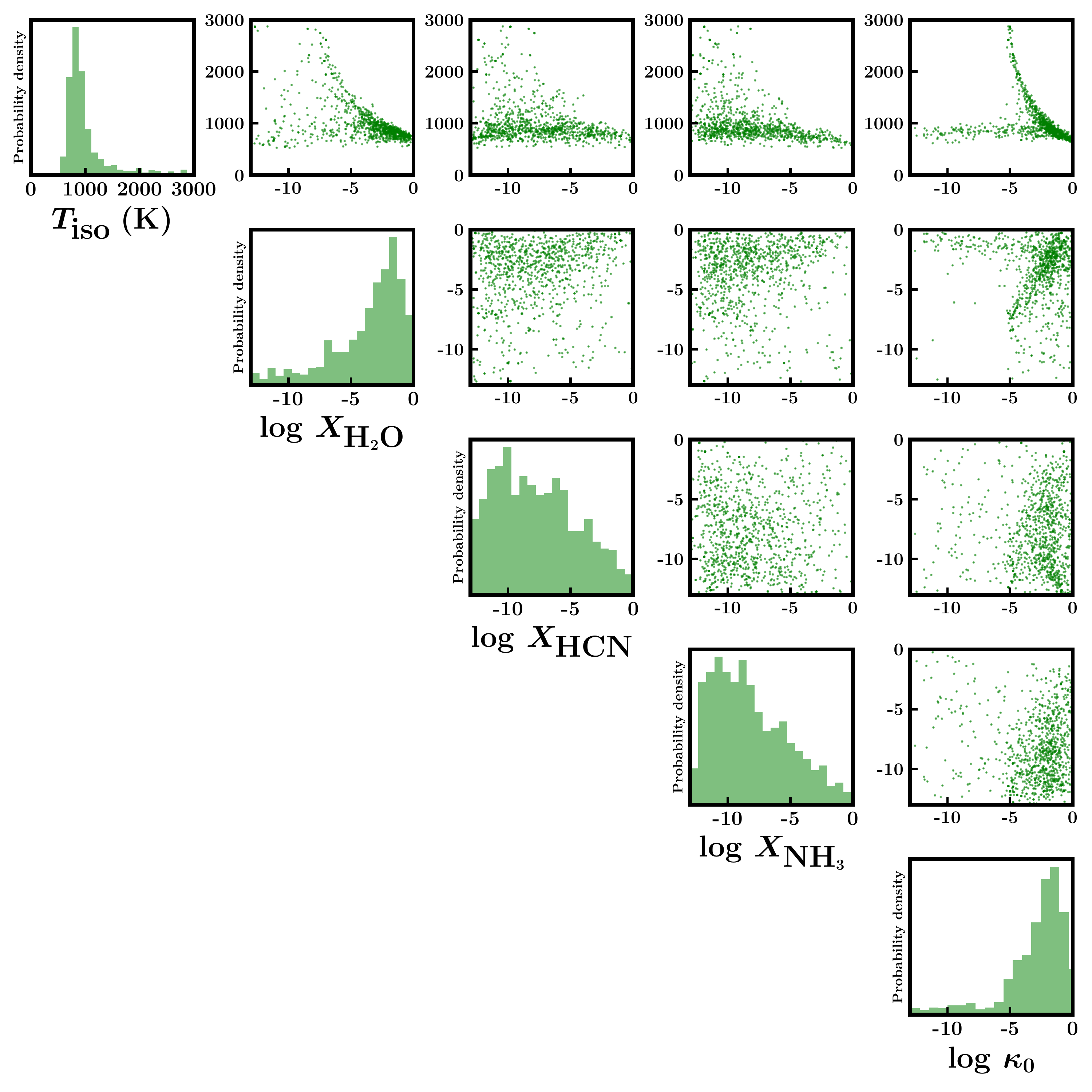}

	\def\arraystretch{1.5}
	\setlength{\arrayrulewidth}{1.3pt}
	\centering
	\Large{
	\vspace{-5.2cm}\hspace{-6.8cm}\begin{tabular}{cc}
		\hline
		Parameter & Value $^{+1\sigma}_{-1\sigma}$ \\
		\hline
		$T_{\textnormal{iso}}$ (K) & $877^{+345}_{-142}$ \\		
		$\log X_{\textnormal{H}_2\textnormal{O}}$ & $-2.84^{+1.72}_{-3.73}$ \\
		$\log X_{\textnormal{HCN}}$ & $-8.03^{+4.03}_{-3.11}$ \\
		$\log X_{\textnormal{NH}_3}$ & $-8.71^{+3.77}_{-2.54}$ \\
		$\log \kappa_0$ & $-2.15^{+1.36}_{-2.58}$ \\
		\hline
	\end{tabular}}
	\vspace{0.5cm}

    \caption{Results of Random Forest retrieval of the WFC3 transmission spectrum of WASP-12b following the methods of \citetalias{PMN18}. While the parameter estimates are consistent within $1\sigma$ with those in Figure \ref{fig:pmn_ns}, the posterior distributions have important differences in shape.}
    \label{fig:pmn_rf}
\end{figure*}

We train 1000 estimators on the training set with a minimum impurity decrease tolerance of 0.01. To begin the training phase, we bin each of the spectra in the training set to the resolution of the WFC3 spectrum, and add random Gaussian noise with a mean of 50 parts per million to each spectral data point. In order to improve the robustness of the predictions, each estimator is shown only 4 of the 13 spectral data points in each training sample. Figure \ref{fig:pmn_rf} shows the distributions of the estimators' predicted parameter values for the WASP-12b spectrum, displaying a close match to figure 1 of \citetalias{PMN18}.

It is important to note that there are some discrepancies between the distributions shown in Figures \ref{fig:pmn_ns} and \ref{fig:pmn_rf}. Most notably, the posterior distributions of H$_2$O abundance and of $\kappa_0$ have broad tails in the Random Forest retrieval which are not found in the Nested Sampling retrieval. These differences arise because the the distributions shown in Figure \ref{fig:pmn_rf} are not true posterior distributions in the Bayesian sense; they are instead the relative densities of the predictions made by 1000 different estimators, some of which perform better than others by design. This means that this method does not necessarily capture the true shape of the posterior distributions and therefore cannot provide a robust estimate of the uncertainties of the predicted parameter values.

\subsection{Extension of Random Forest method} \label{sub:extension}

The differences between the shapes of the posterior distributions produced by the two different retrieval methods motivate the development of a new method, still employing machine learning in the form of the Random Forest algorithm, but yielding results that capture the uncertainty in parameter estimates more accurately. A diagram depicting this new approach is shown in Figure \ref{fig:rf_flowchart}. We begin by producing a training data set in the same way as before, but we do not add noise to the model spectra. Before the training phase, we normalise the parameter values in the training data set so that they all lie between 0 and 1. This ensures that the loss function does not favour any one parameter over another. 

Once the estimators have been trained on this noise-free data set and used to predict parameter values, the likelihoods of those predictions are calculated by comparing the observed spectrum to a forward model produced with the predicted parameter values (see equation \ref{eq:like}). This set of predictions and associated likelihoods serves as the likelihood function for the retrieval, allowing the marginalised posterior for each parameter and pair of parameters to be computed. By calculating the likelihood of each prediction, the algorithm should no longer produce long tails that are not found in a Nested Sampling retrieval, since these predictions will have lower likelihoods and will be penalised accordingly. Since we impose a Gaussian likelihood, this method differs from other machine learning-based approaches to retrievals, which are typically likelihood-free.

The procedure initially trains a set of 1000 estimators to compute an initial posterior estimate. However, in order to ensure that enough estimators have been trained to sample the parameter space thoroughly, more estimators are added in batches of 1000 until the symmetric Kullback-Leibler divergence ($\Delta_{\rm KL}$) between successive posterior distributions falls below a certain tolerance. $\Delta_{\rm KL}$ is defined as

\begin{equation}
\begin{split}
    \Delta_{\rm KL} =& \sum_x \bigg[ p_i (\theta = x|d) \log \bigg( \frac{p_i (\theta = x|d)}{p_{i+1} (\theta = x|d)} \bigg) \\ &+ p_{i+1} (\theta = x|d) \log \bigg( \frac{p_{i+1} (\theta = x|d)}{p_{i} (\theta = x|d)} \bigg) \bigg]
\end{split}
\end{equation}
where $p_i (\theta = x|d)$ is the posterior probability that $\theta = x$ given by a forest of $(1000 \times i)$ trees. Figure \ref{fig:rf_flowchart} shows a flowchart describing our extended Random Forest retrieval method.

\begin{figure*}
\includegraphics[width=0.7\textwidth]{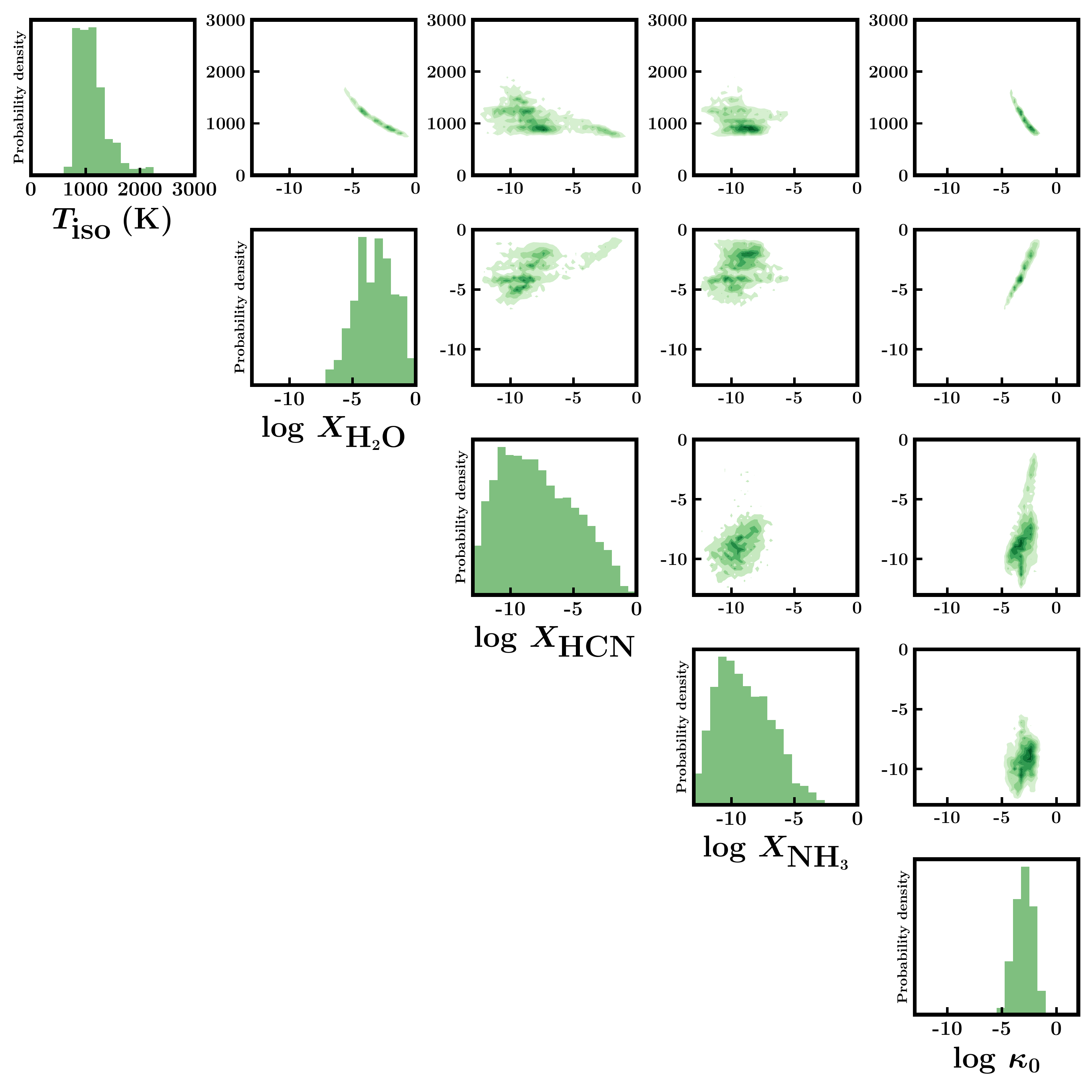}

	\def\arraystretch{1.5}
	\setlength{\arrayrulewidth}{1.3pt}
	\centering
	\Large{
	\vspace{-5.2cm}\hspace{-6.8cm}\begin{tabular}{cc}
		\hline
		Parameter & Value $^{+1\sigma}_{-1\sigma}$ \\
		\hline
		$T_{\textnormal{iso}}$ (K) & $1036^{+345}_{-186}$ \\		
		$\log X_{\textnormal{H}_2\textnormal{O}}$ & $-3.13^{+1.34}_{-1.75}$ \\
		$\log X_{\textnormal{HCN}}$ & $-8.49^{+3.67}_{-2.78}$ \\
		$\log X_{\textnormal{NH}_3}$ & $-9.53^{+2.59}_{-1.97}$ \\
		$\log \kappa_0$ & $-2.96^{+0.71}_{-0.92}$ \\
		\hline
	\end{tabular}}
	\vspace{0.5cm}

    \caption{Results of extended Random Forest retrieval of the WFC3 transmission spectrum of WASP-12b, using the same forward model as in \citetalias{PMN18}. The parameter estimates and posterior distributions provide a better match to the Nested Sampling retrieval shown in Figure \ref{fig:pmn_ns}, compared to Figure \ref{fig:pmn_rf}.}
    \label{fig:pmn_rf_plus}
\end{figure*}

Figure \ref{fig:pmn_rf_plus} shows the posterior distribution from a retrieval of the same WASP-12b spectrum using the extended Random Forest method as described above. The retrieved parameter values are consistent within 1$\sigma$ with those obtained in the previous two retrievals. However, the main difference between the results of this retrieval and those of the previous Random Forest retrieval is that the shape of the posterior distribution matches the Nested Sampling posterior more closely. The extended tails found in Figure \ref{fig:pmn_rf} are no longer present. This is reflected in the reported uncertainties in the parameter estimates. Whereas the Random Forest retrieval following the approach from \citetalias{PMN18} gives lower bounds far below those given by the Nested Sampling retrieval, the extended approach gives error bounds that are in line with the Nested Sampling result.

In order to obtain a good sampling of the parameter space for this problem using the extended Random Forest approach, a much larger number of estimators is required than the ensemble of 1000 used in the method outlined in \citetalias{PMN18}. Convergence is reached after 17000 estimators to produce the posterior distributions shown in Figure \ref{fig:pmn_rf_plus}, with a mean tree depth of 39. A reasonable result can be produced using a higher tolerance which converges after approximately 10000 estimators have been generated. Simply increasing the number of trees without the likelihood evaluation step is not sufficient to obtain a more accurate retrieval; this is discussed in more detail in Section \ref{section:HD}. While this increases the computational cost of the approach somewhat, the longest step in the retrieval is still the generation of the training data, since the forward model must be run 100000 times to create the full training set.

We investigated the effect of lowering the amount of training data to reduce computation time. This would reduce both the time taken for the training set to be produced and the training time itself, since the Random Forest algorithm trains more quickly on a smaller data set. Decreasing the amount of training data from 100000 spectra to 50000 yields resulting posterior distributions that are not well-sampled. We conclude that in this case a significant reduction in the amount of data used to train the estimators is not feasible.

\section{Applications} \label{section:HD}

Having demonstrated that we can reproduce the results of \citetalias{PMN18}, and having extended their method to produce a result closer to that of a Nested Sampling retrieval, we now compare our new approach to a current state-of-the-art retrieval framework that uses a fully numerical forward model. We no longer use the forward model from \citet{HK17}, instead adopting the modelling paradigm from AURA \citep{pinhas_18}, a retrieval framework that has been validated against synthetic spectra and  used to retrieve atmospheric properties, including H$_2$O abundances, for a range of planets \citep[e.g.,][]{Pinhas_19,Welbanks_19b}.

\subsection{Validation} \label{sub:val}

\begin{figure*}
\includegraphics[width=\linewidth]{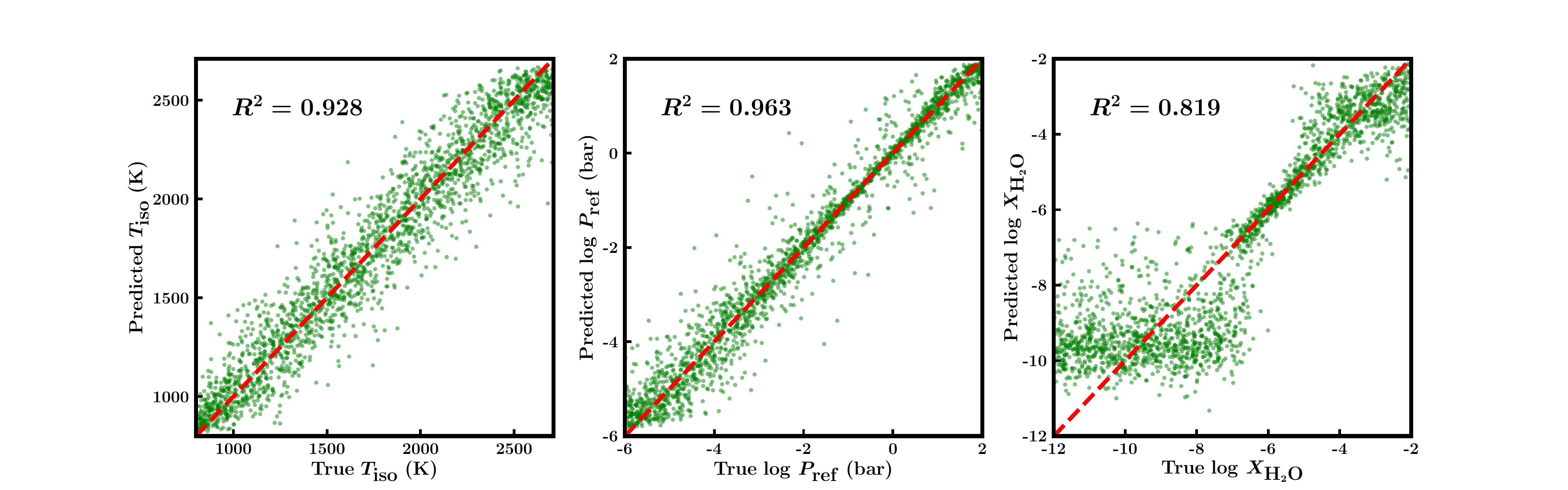}
    \caption{True values versus Random Forest predictions for each parameter in the forward model, for a test set of 2000 synthetic models. The coefficient of determination ($R^2$) indicates the correlation between the true and predicted values, with $R^2$ close to 1 implying a strong correlation.}
    \label{fig:synthetic_test}
\end{figure*}

\begin{figure}
\includegraphics[width=\columnwidth]{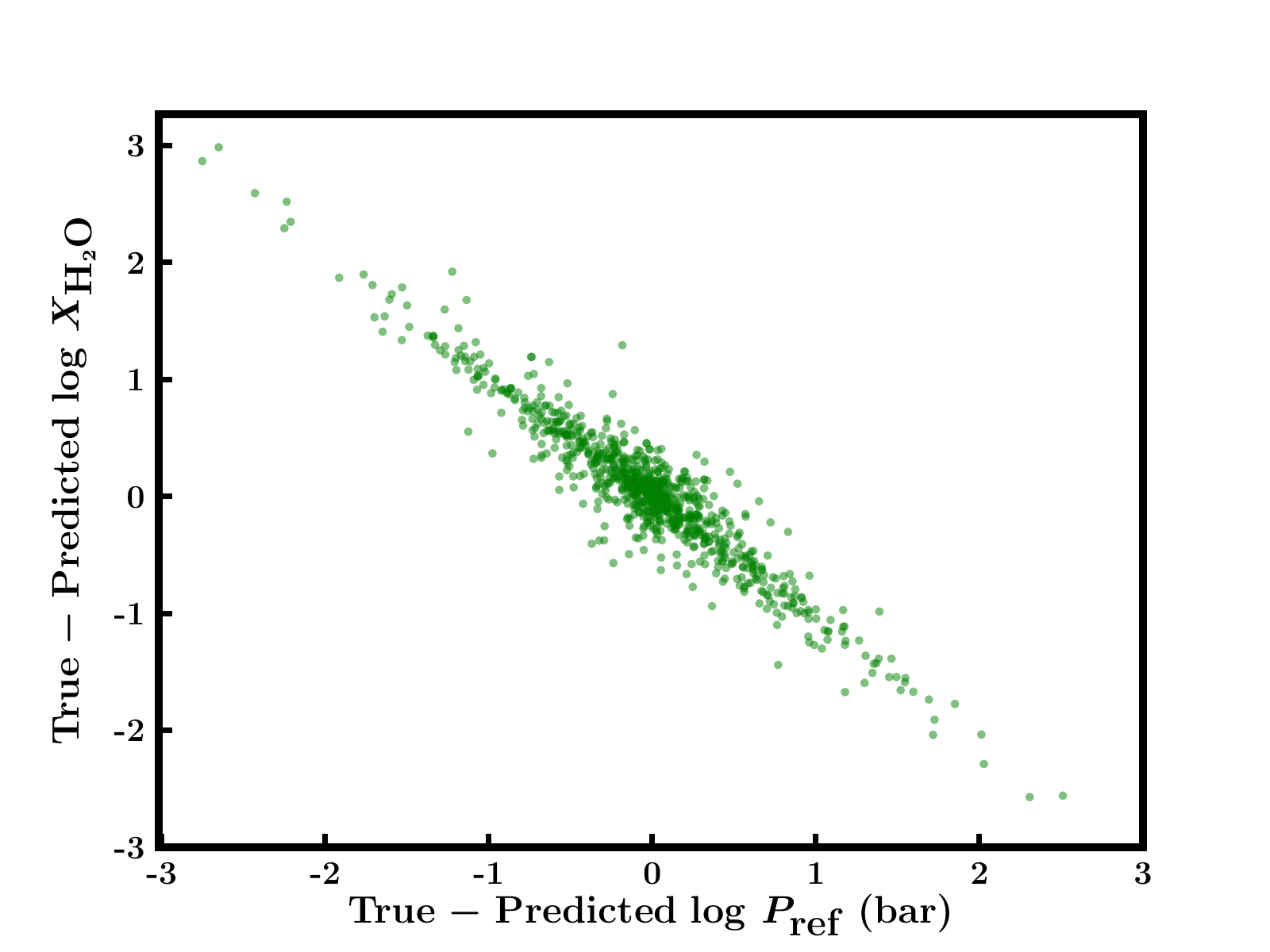}
    \caption{Difference between true and predicted values of $P_{\rm ref}$ and $X_{\textnormal{H}_2\textnormal{O}}$ for synthetic models with $\log X_{\textnormal{H}_2\textnormal{O}} \geq -6$. We find that predictions of $P_{\rm ref}$ that are above the true value correspond to predictions of $X_{\textnormal{H}_2\textnormal{O}}$ below the true value and vice-versa.}
    \label{fig:degeneracy}
\end{figure}

We begin by demonstrating our algorithm's ability to accurately estimate parameter values from synthetic spectra. The AURA forward model is used to generate synthetic spectra in the wavelength range of WFC3 for training and testing our algorithm, assuming a cloud-free atmosphere and an isothermal temperature profile. Since WFC3 transmission spectra only provide nominal constraints on the temperature structure of the atmosphere \citep{Barstow_13}, assuming an isothermal temperature profile has little effect on retrievals with present data and is sufficient for the purposes of this study. Cloud properties are not considered since they are difficult to constrain without including data from optical wavelengths. 

The model atmosphere is divided into 100 pressure layers, which are evenly log-spaced from $10^{-6} - 10^2$ bar. The main opacity sources in the model are H$_2$O and collision-induced absorption (CIA) due to H$_2$-H$_2$ and H$_2$-He. The cross-sections for these opacity sources are computed by \citet{sid_17} using line lists from the HITEMP database for H$_2$O \citep{HITEMP} and the HITRAN database for CIA \citep{Richard_12}. We fix $R_p$ to $1.41R_J$ and leave the reference pressure $P_{\textnormal{ref}}$ as a free parameter to be retrieved. It was demonstrated by \citet{Welbanks_19} that fixing one of the planetary radius and reference pressure and retrieving the other does not affect the retrieved values of the other parameters, and so we arbitrarily choose the retrieved radius of HD~209458b from Case 3 of that paper. The model therefore has 3 free parameters: $T_{\textnormal{iso}}$, $P_{\textnormal{ref}}$ and the water abundance $X_{\textnormal{H}_2\textnormal{O}}$. This choice of paramaterisation is appropriate for WFC3 spectra of hot Jupiters, which to date have been found to be most sensitive to planetary radius, temperature and H$_2$O abundance \citep{Tsiaras_18,Welbanks_19}. The prior ranges for each parameter are shown in Table \ref{tab:priors_hd}.

We use this forward model to produce a training data set of 8000 model spectra and a validation data set of 2000 spectra. Each data set has randomly generated parameters, and no spectra from the validation set appear in the training set. The spectra are produced at a high resolution ($R=1000$) in the wavelength range of WFC3 ($1.1-1.7$ $\mu$m). Experimenting with larger and smaller training data sets suggests that this is the minimum size for the algorithm to be able to accurately learn the relationship between the input spectrum and the output parameters. The Random Forest is set up in the same manner as before, using the same hyperparameters to train on normalised data. The trained Random Forest is then used to predict the parameter values for the 2000 synthetic spectra in the test set. As described in Section \ref{section:methods}, we evaluate the likelihoods of the predictions made by each tree and take the median of the corresponding posterior distribution to be the predicted parameter value.

Figure \ref{fig:synthetic_test} shows the outcome of our method when applied to these 2000 synthetic spectra. The $R^2$ coefficient of determination is close to unity for each parameter, suggesting that the retrieval is able to recover the input parameters well. For models with $\log X_{\textnormal{H}_2\textnormal{O}} \lesssim -6$ the correlation is much lower; this is to be expected since these cases correspond to non-detections and is consistent with the findings of \citetalias{PMN18}. For models with $\log X_{\textnormal{H}_2\textnormal{O}} \lesssim -6$, the spread in the results is caused by a degeneracy between $P_{\rm ref}$ and $X_{\textnormal{H}_2\textnormal{O}}$, as shown in Figure \ref{fig:degeneracy}; it is possible to fit the same spectrum by increasing $P_{\rm ref}$ and decreasing $X_{\textnormal{H}_2\textnormal{O}}$. This degeneracy has been found previously when analysing WFC3 spectra \citep[e.g.,][]{Pinhas_19,Welbanks_19}.

\subsection{Retrieval of WFC3 transmission spectrum}

Having validated our retrieval method using synthetic data, we now apply the algorithm to a real data set for direct comparison against an AURA retrieval. We consider the observed WFC3 transmission spectrum of the hot Jupiter HD 209458b \citep{Deming_13}, which consists of 29 data points in the spectral range $1.1-1.7$ $\mu$m. We use this planet as a representative case to validate our method since it is one of the most well-studied planets in the literature, with high quality spectral data available. Additionally, the transmission spectrum of this planet has recently been analysed in numerous retrieval studies, \citep[e.g.,][]{Barstow_17,MacDonald_17,Welbanks_19}.

We first carry out a Nested Sampling retrieval using the same parameterisation as in Section \ref{sub:val}. For this retrieval, the model is initially evaluated at a higher resolution, covering 1000 wavelength points from $1.1-1.7$ $\mu$m. The model spectrum is subsequently convolved with the WFC3 point spread function, integrated over the instrument function and binned to the resolution of the observed data. The binned spectrum is used to evaluate the likelihood function for each model (see equation \ref{eq:like}). Further detail on this retrieval approach and validation using synthetic data can be found in \citet{pinhas_18}. We find that a Nested Sampling retrieval setup using 1000 live points is sufficient to obtain robust parameter estimates.

\begin{figure}
\includegraphics[width=\columnwidth]{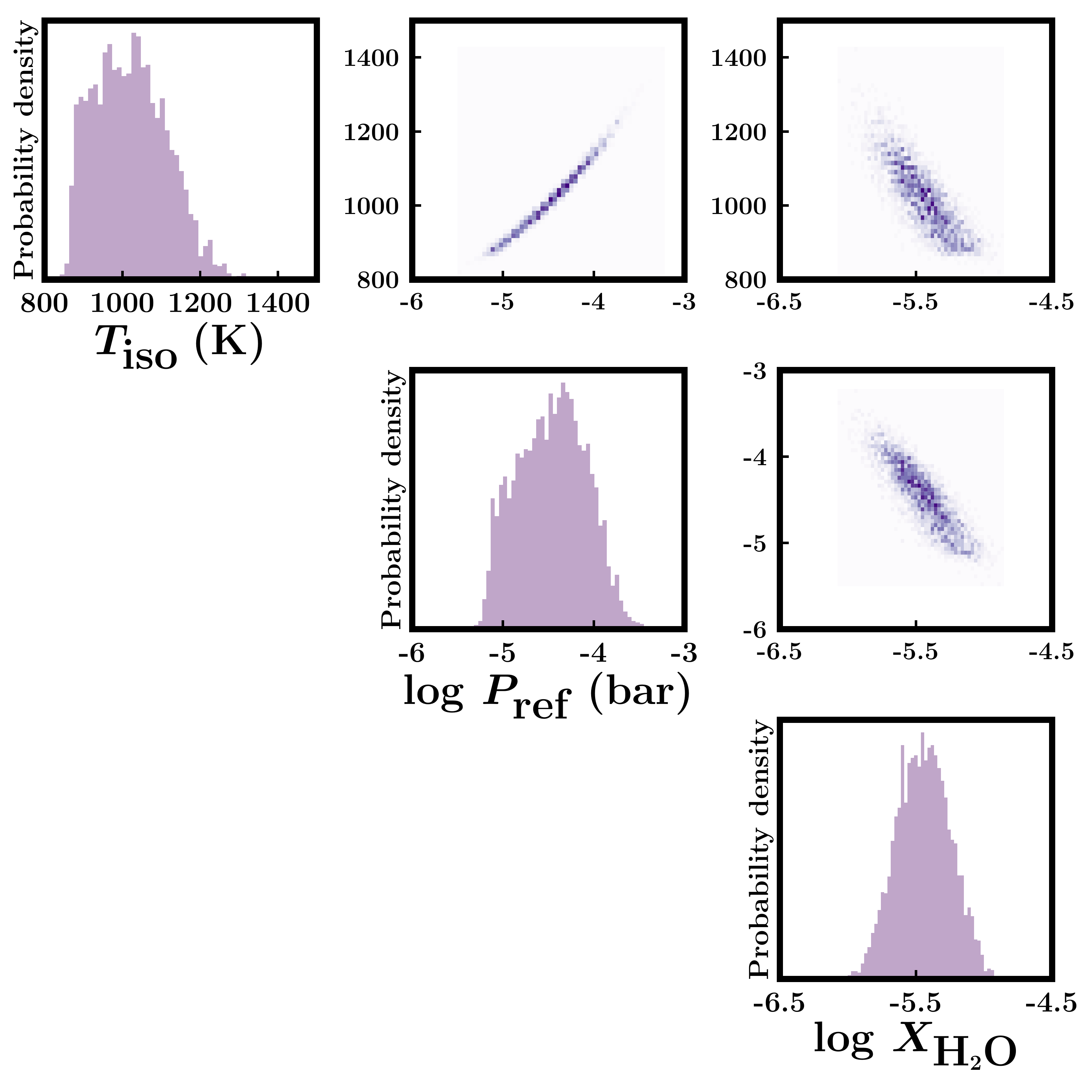}

	\def\arraystretch{1.5}
	\setlength{\arrayrulewidth}{1.3pt}
	\centering
	\large{
	\vspace{-2.82cm}\hspace{-2.8cm}\begin{tabular}{cc}
		\hline
		Parameter & Value $^{+1\sigma}_{-1\sigma}$ \\
		\hline
		$T_{\textnormal{iso}}$ (K) & $1017^{+102}_{-98}$ \\
		$\log \, P_{\textnormal{ref}}$ (bar) & $-4.45^{+0.39}_{-0.44}$ \\
		$\log X_{\textnormal{H}_2\textnormal{O}}$ & $-5.44 \pm 0.20$ \\
		\hline
	\end{tabular}}
	\vspace{0.5cm}

    \caption{Results of Nested Sampling retrieval of the WFC3 transmission spectrum of HD 209458b.}
    \label{fig:ns_hd_3param}
\end{figure}

The retrieved posterior distributions, abundance estimates and uncertainties are displayed in Figure \ref{fig:ns_hd_3param}. These compare very closely to the results from Case 3 of \citet{Welbanks_19}. The only notable difference between the two retrievals is that our results show smaller error bars for the estimated reference pressure $P_{\textnormal{ref}}$. This can be attributed to fixing $R_p$ rather than retrieving it. We again find a degeneracy between $P_{\rm ref}$ and $X_{\textnormal{H}_2\textnormal{O}}$, in agreement with our findings from the model validation.

\begin{table}
	\centering
	\caption{Description of priors for retrievals of the WFC3 transmission spectrum of HD 209458b.}
	\hfill \\
	\label{tab:priors_hd}
	\def\arraystretch{1.5}
	\setlength{\arrayrulewidth}{1.3pt}
	\begin{tabular}{cccc}
		\hline
		Parameter & Lower Bound & Upper Bound & Prior \\
		\hline
		$T_{\textnormal{iso}}$ (K) & $700$ & $2810$ & uniform \\
		$P_{\textnormal{\rm ref}}$ (bar) & $10^{-6}$ & $10^{2}$ & log-uniform \\
		$X_{\textnormal{H}_2\textnormal{O}}$ & $10^{-12}$ & $10^{-2}$ & log-uniform\\
		\hline
	\end{tabular}
\end{table}

\begin{figure}
\includegraphics[width=\columnwidth]{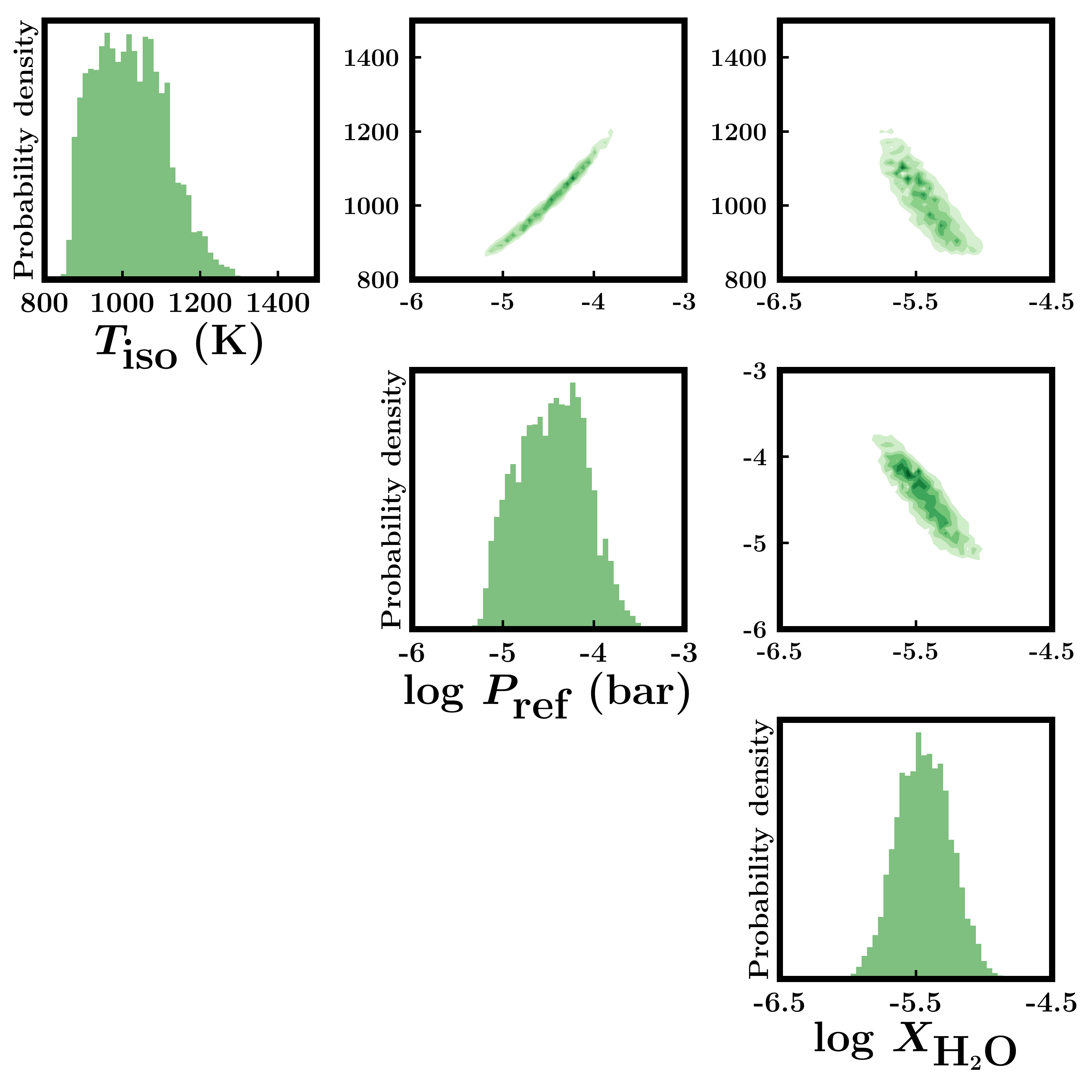}

	\def\arraystretch{1.5}
	\setlength{\arrayrulewidth}{1.3pt}
	\centering
	\large{
	\vspace{-2.82cm}\hspace{-2.8cm}\begin{tabular}{cc}
		\hline
		Parameter & Value $^{+1\sigma}_{-1\sigma}$ \\
		\hline
		$T_{\textnormal{iso}}$ (K) & $1009^{+99}_{-94}$ \\
		$\log \, P_{\textnormal{ref}}$ (bar) & $-4.48^{+0.37}_{-0.43}$ \\
		$\log X_{\textnormal{H}_2\textnormal{O}}$ & $-5.47 \pm 0.20$ \\
		\hline
	\end{tabular}}
	\vspace{0.5cm}
	
    \caption{Results of extended Random Forest retrieval of the WFC3 transmission spectrum of HD 209458b. The results are in agreement with the Nested Sampling retrieval in Figure \ref{fig:ns_hd_3param}.}
    \label{fig:rf_hd_3param}
\end{figure}

We attempt to reproduce the results of this Nested Sampling retrieval using the extended Random Forest method as described in Section \ref{sub:extension}. For this case study we use the same training set of 8000 spectra from Section \ref{sub:val}. The Random Forest is set up in the same manner as before, using the same hyperparameters to train on normalised data. The likelihood function for each estimator is evaluated every time 1000 more estimators have been trained, and we find that the posterior distribution converges once 12000 estimators have been produced. Marginalised posterior distributions are then created from this likelihood function to obtain the results, which are displayed in Figure \ref{fig:rf_hd_3param}.

\begin{figure*}
\includegraphics[width=\textwidth,trim={0 1.6cm 0 0},clip]{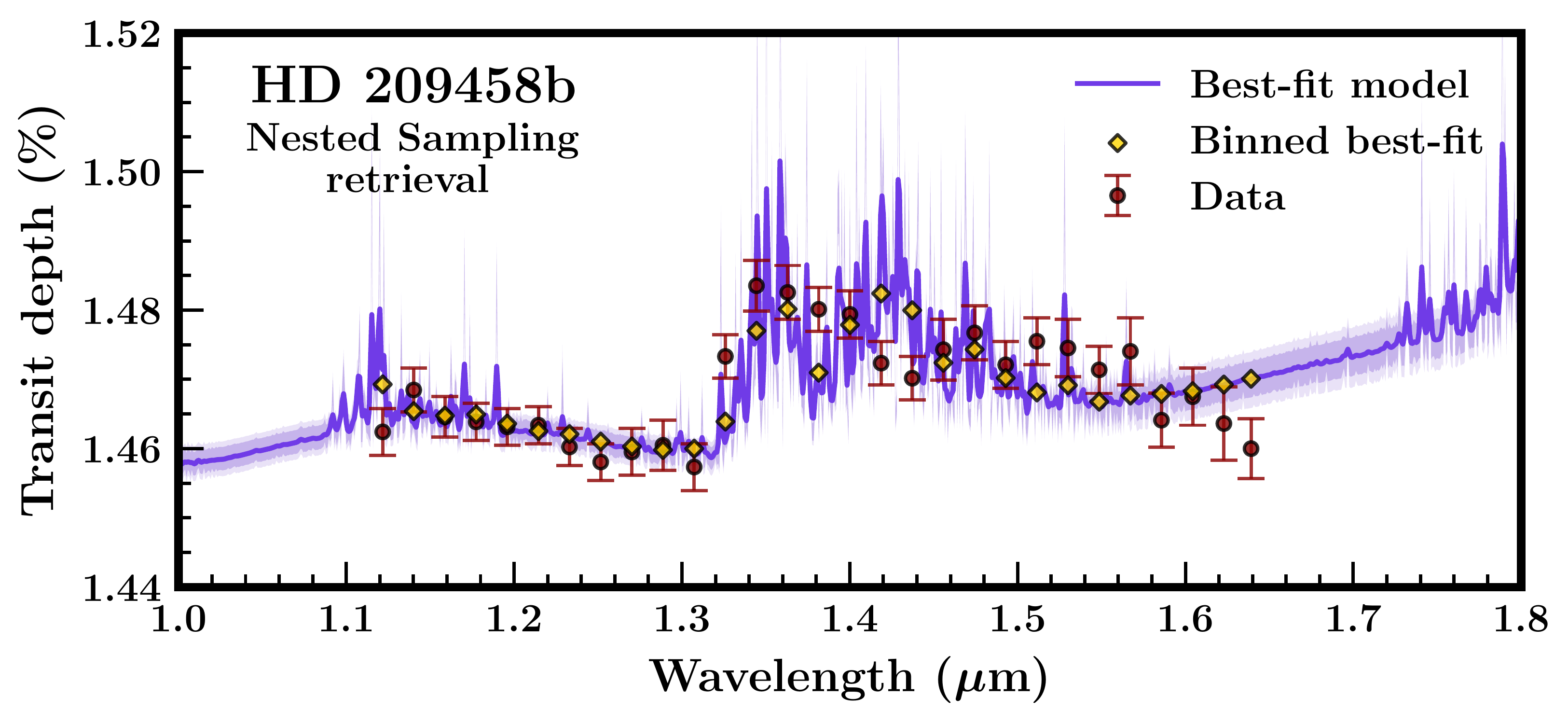}
\includegraphics[width=\textwidth]{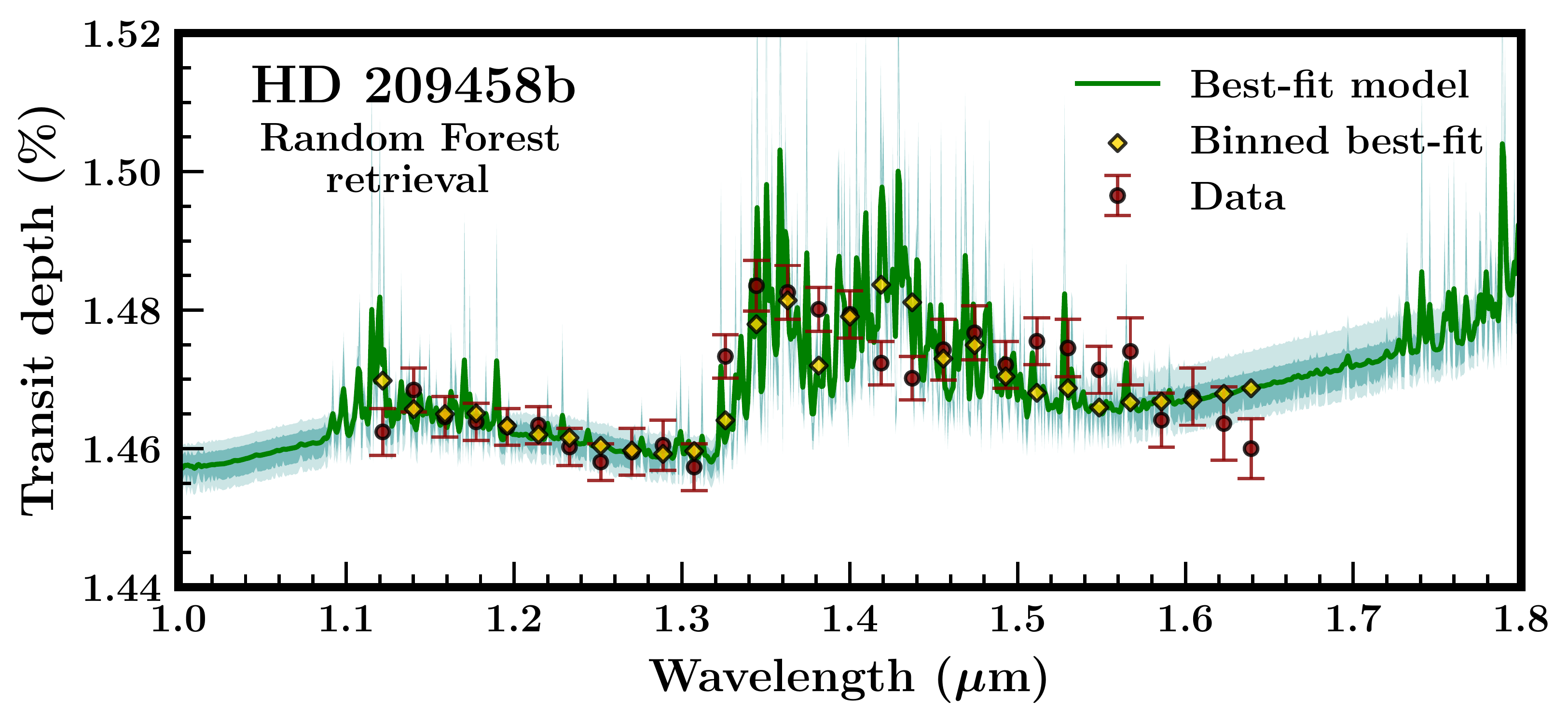}
    \caption{\textit{Top:} Best-fitting model spectrum from the Nested Sampling retrieval of the HD 209458b WFC3 transmission spectrum. \textit{Bottom:} Best-fitting model spectrum from the extended Random Forest retrieval of the same spectrum. The shaded regions represent the $1\sigma$ and $2\sigma$ contours, produced by drawing 1000 spectra from the posterior distributions from each retrieval. The best-fitting spectra have been smoothed with a Gaussian filter for clarity.}
    \label{fig:hd_3param_spec}
\end{figure*}

In this case, the extended Random Forest retrieval produces extremely similar results to the Nested Sampling retrieval. The best-fitting model spectra, along with 1$\sigma$ and 2$\sigma$ uncertainties, from the two retrievals are shown in Figure \ref{fig:hd_3param_spec}. The parameter estimates and uncertainties are directly compared in Table \ref{tab:ret_vals_3param}; both the retrieved median values and the 1$\sigma$ uncertainties are almost identical. It is also clear from the joint distributions shown in Figure \ref{fig:rf_hd_3param} that the extended Random Forest retrieval has found the same degeneracies between parameters as the Nested Sampling retrieval. This result demonstrates for the first time that, for a given observation instance, a machine learning-based approach to atmospheric retrieval can not only obtain similar parameter estimates to a traditional retrieval, but that it can also deal with uncertainties and degeneracies in a robust and accurate way.

\begin{table}
	\centering
	\caption{Comparison of Nested Sampling (NS) and extended Random Forest (RF+) retrieved parameter values from the WFC3 transmission spectrum of HD 209458b.}
	\hfill \\
	\label{tab:ret_vals_3param}
	\def\arraystretch{1.5}
	\setlength{\arrayrulewidth}{1.3pt}
	\large{
	\begin{tabular}{ccc}
		\hline
		Parameter & NS Value $^{+1\sigma}_{-1\sigma}$ & RF+ Value $^{+1\sigma}_{-1\sigma}$\\
		\hline
		$T_{\textnormal{iso}}$ (K) & $1017^{+102}_{-98}$ & $1009^{+99}_{-94}$ \\
		$\log \, P_{\textnormal{ref}}$ (bar) & $-4.45^{+0.39}_{-0.44}$ & $-4.48^{+0.37}_{-0.43}$ \\
		$\log X_{\textnormal{H}_2\textnormal{O}}$ & $-5.44 \pm 0.20$ & $-5.47 \pm 0.20$ \\
		\hline
	\end{tabular}}
\end{table}

\subsection{Addition of unconstrained free parameter}

The retrieval analysis of the WASP-12b transmission spectrum from \citetalias{PMN18} has also been reproduced in \citet{Cobb_19}. In that paper they find that in certain cases, the Random Forest retrieval can sometimes return a narrow posterior for a free parameter that should not be constrained; they demonstrate this by finding a synthetic spectrum following the model from \citet{HK17} for which the Random Forest confidently predicts H$_2$O, HCN and NH$_3$ abundances that are not the true values used to generate the model. In the present paper we investigate this issue further in order to determine whether the \citetalias{PMN18} Random Forest approach might incorrectly infer certain parameter values in cases where a traditional retrieval would (correctly) not be able to constrain that value. We also consider whether a similar problem would occur if our extended Random Forest method was used instead.

\begin{figure}
\includegraphics[width=\columnwidth]{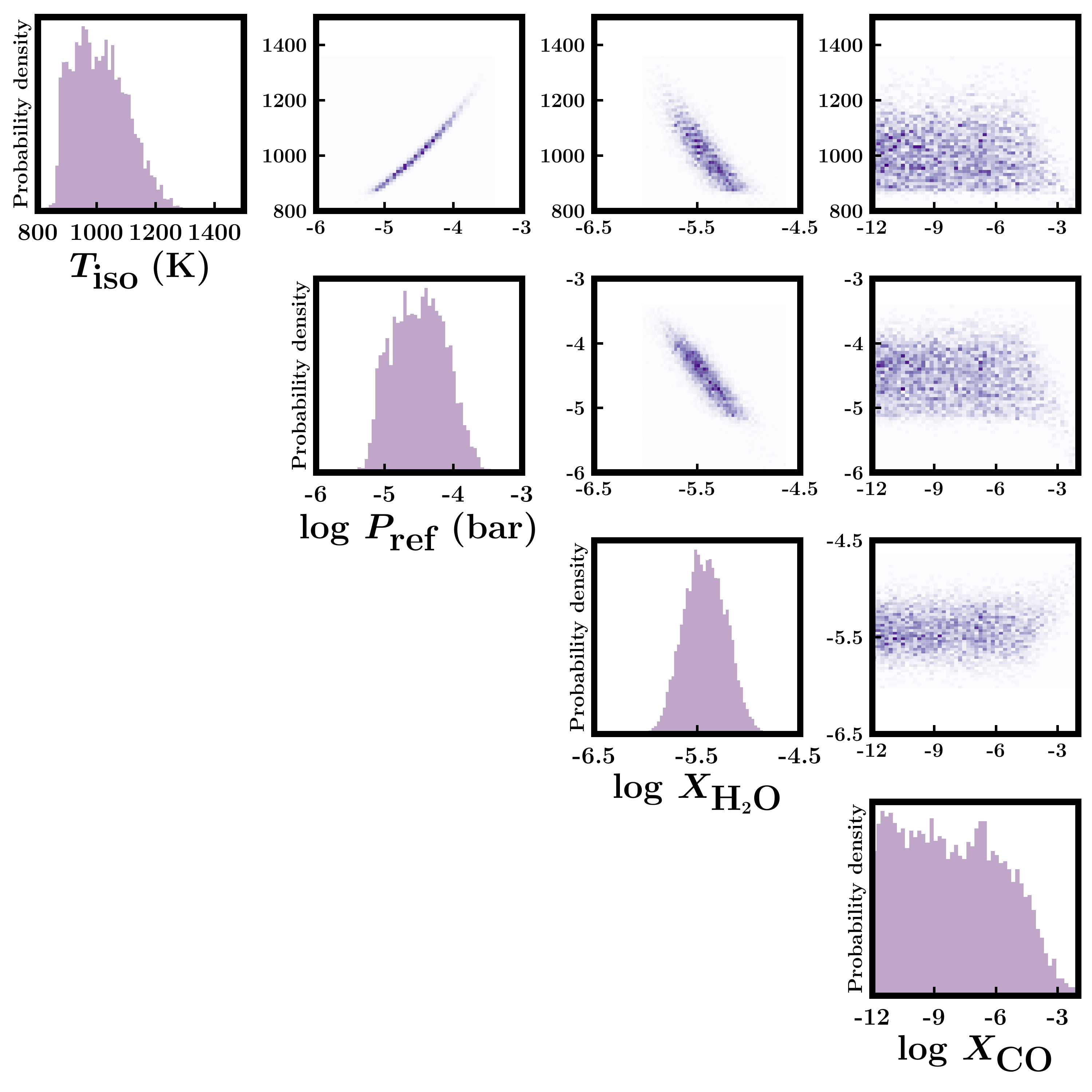}

	\def\arraystretch{1.5}
	\setlength{\arrayrulewidth}{1.3pt}
	\centering
	\large{
	\vspace{-3.4cm}\hspace{-4.0cm}\begin{tabular}{cc}
		\hline
		Parameter & Value $^{+1\sigma}_{-1\sigma}$ \\
		\hline
		$T_{\textnormal{iso}}$ (K) & $1004^{+106}_{-91}$ \\
		$\log \, P_{\textnormal{ref}}$ (bar) & $-4.50^{+0.40}_{-0.44}$ \\
		$\log X_{\textnormal{H}_2\textnormal{O}}$ & $-5.43^{+0.22}_{-0.21}$ \\
		$\log X_{\textnormal{CO}}$ & $-8.22^{+2.86}_{-2.64}$ \\
		\hline
	\end{tabular}}
	\vspace{0.5cm}

    \caption{Results of Nested Sampling retrieval of the WFC3 transmission spectrum of HD 209458b, including CO abundance as a free parameter. The non-detection of CO is consistent with expectation, given the weak CO features in the spectral range of the data.}
    \label{fig:ns_hd_4param}
\end{figure}

In order to highlight this issue, we carry out a second set of retrievals of the WFC3 spectrum of HD~209458b, but this time including CO abundance as a free parameter in the forward model, with a log-uniform prior ranging from $10^{-12}-10^{-2}$. We choose to add CO since this molecule does not have strong features in the spectral range of the data. Previous studies such as \citet{Welbanks_19} have therefore been unable to constrain the CO abundance from this spectrum. We verify this by first carrying out a Nested Sampling retrieval of the spectrum, whose results can be seen in Figure \ref{fig:ns_hd_4param}. As expected, the estimated values of $T_{\textnormal{iso}}, P_{\textnormal{ref}}$ and $\log X_{\textnormal{H}_2\textnormal{O}}$ remain very close to those from the retrieval that did not include CO (see Figure \ref{fig:ns_hd_3param}), but the CO abundance itself is not unconstrained. This setup should therefore provide a test of the capabilities of both the \citetalias{PMN18} and our Random Forest retrieval methods to deal with an unconstrained free parameter in the model.

\begin{figure}
\includegraphics[width=\columnwidth]{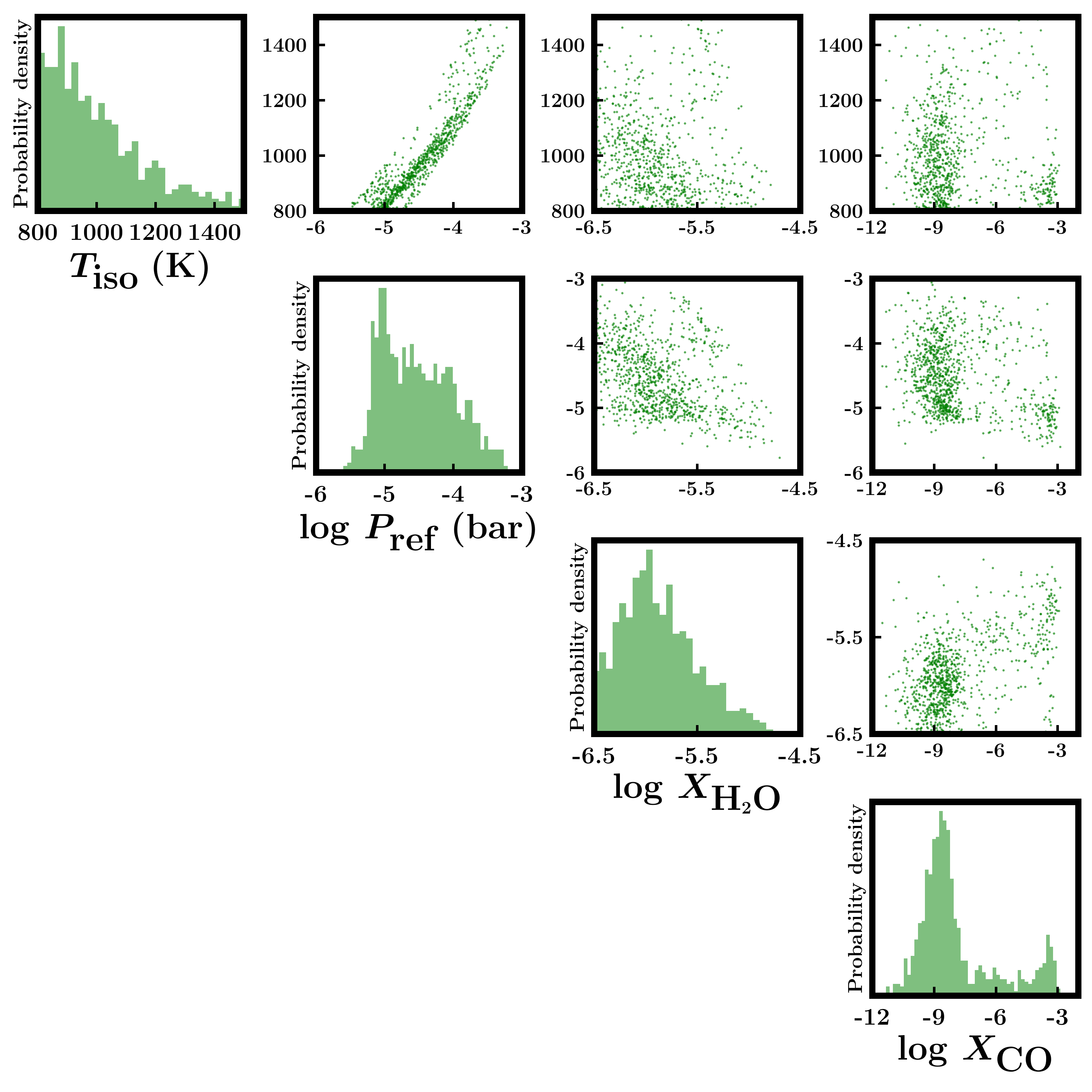}

	\def\arraystretch{1.5}
	\setlength{\arrayrulewidth}{1.3pt}
	\centering
	\large{
	\vspace{-3.4cm}\hspace{-4.0cm}\begin{tabular}{cc}
		\hline
		Parameter & Value $^{+1\sigma}_{-1\sigma}$ \\
		\hline
		$T_{\textnormal{iso}}$ (K) & $935^{+222}_{-124}$ \\
		$\log \, P_{\textnormal{ref}}$ (bar) & $-4.58^{+0.64}_{-0.48}$ \\
		$\log X_{\textnormal{H}_2\textnormal{O}}$ & $-5.95^{+0.44}_{-0.37}$ \\
		$\log X_{\textnormal{CO}}$ & $-8.50^{+2.81}_{-0.90}$ \\
		\hline
	\end{tabular}}
	\vspace{0.5cm}

    \caption{Results of Random Forest retrieval of the WFC3 transmission spectrum of HD 209458b, including CO abundance as a free parameter, following the methods of \citetalias{PMN18}. The peak in the CO posterior is unphysical given that there is no strong CO feature in the spectral range of the data.}
    \label{fig:pmn_rf_hd_4param}
\end{figure}

We begin the machine learning approach by generating a training set consisting of 160000 model spectra, which we use for both the \citetalias{PMN18} and for the extended Random Forest retrievals. We use the same training set in both cases to ensure that the only difference between the two retrievals is the implementation of the algorithm. First we employ the methods of \citetalias{PMN18} to perform the retrieval using this data set. We add Gaussian noise to the training set and train 1000 estimators on the noisy spectra. Histograms of the results along with parameter estimates are shown in Figure \ref{fig:pmn_rf_hd_4param}. The shapes of the temperature and water abundance distributions differ somewhat from the Nested Sampling posteriors, but what is most notable is the apparent peak around $-8.5$ in $\log$ CO abundance which is not present at all in the Nested Sampling case. As in \citet{Cobb_19}, the algorithm is overconfident in its prediction of a parameter value which it should not be able to constrain.

\begin{figure}
\includegraphics[width=\columnwidth]{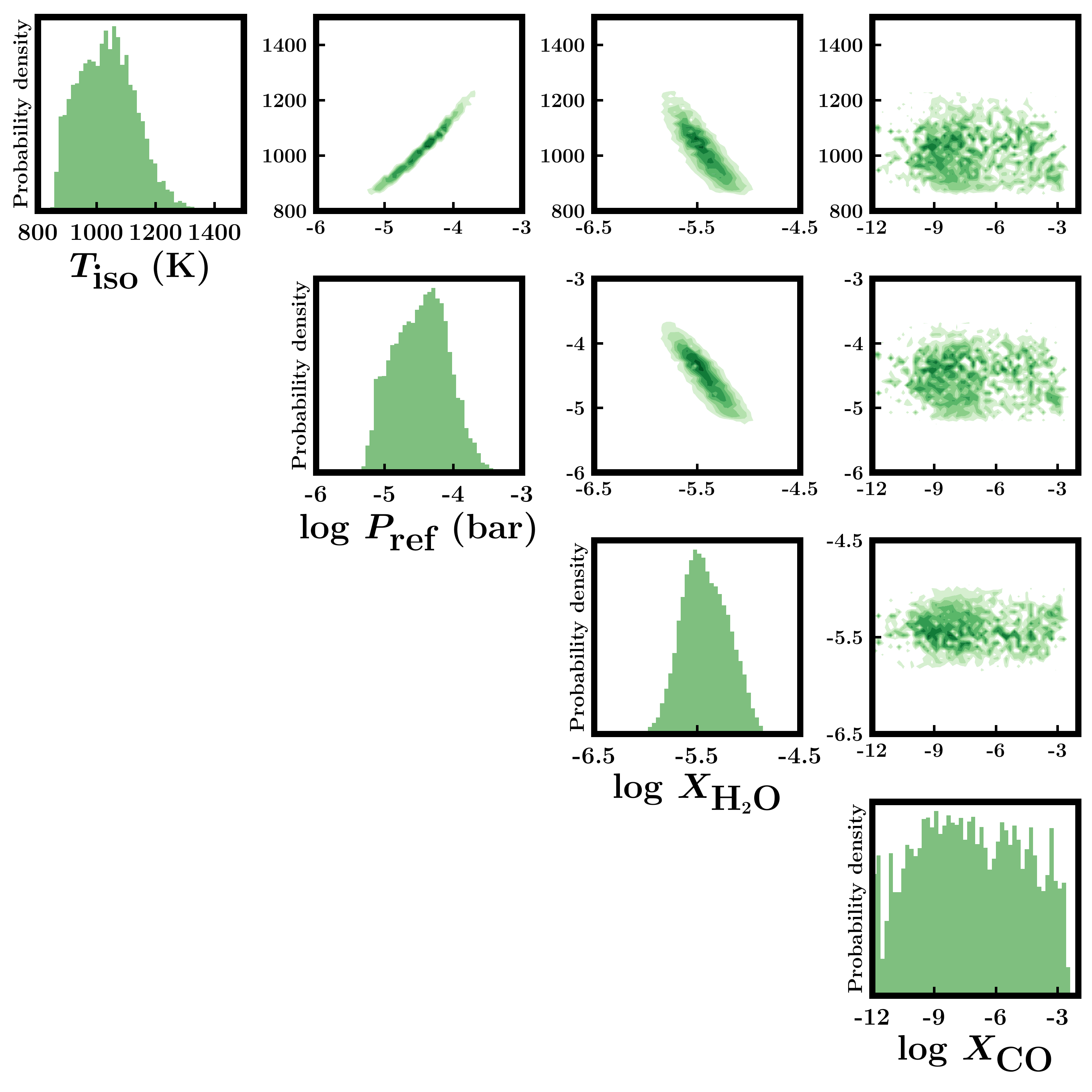}

	\def\arraystretch{1.5}
	\setlength{\arrayrulewidth}{1.3pt}
	\centering
	\large{
	\vspace{-3.4cm}\hspace{-4.0cm}\begin{tabular}{cc}
		\hline
		Parameter & Value $^{+1\sigma}_{-1\sigma}$ \\
		\hline
		$T_{\textnormal{iso}}$ (K) & $1026^{+100}_{-103}$ \\
		$\log \, P_{\textnormal{ref}}$ (bar) & $-4.49^{+0.38}_{-0.45}$ \\
		$\log X_{\textnormal{H}_2\textnormal{O}}$ & $-5.45^{+0.25}_{-0.21}$ \\
		$\log X_{\textnormal{CO}}$ & $-7.40^{+3.00}_{-2.77}$ \\
		\hline
	\end{tabular}}
	\vspace{0.5cm}

    \caption{Results of extended Random Forest retrieval of the WFC3 transmission spectrum of HD 209458b, including CO abundance as a free parameter. The results are consistent with expectations and the Nested Sampling retrieval shown in Figure \ref{fig:ns_hd_4param}.}
    \label{fig:rf_plus_hd_4param}
\end{figure}

\begin{figure*}
\includegraphics[width=\textwidth]{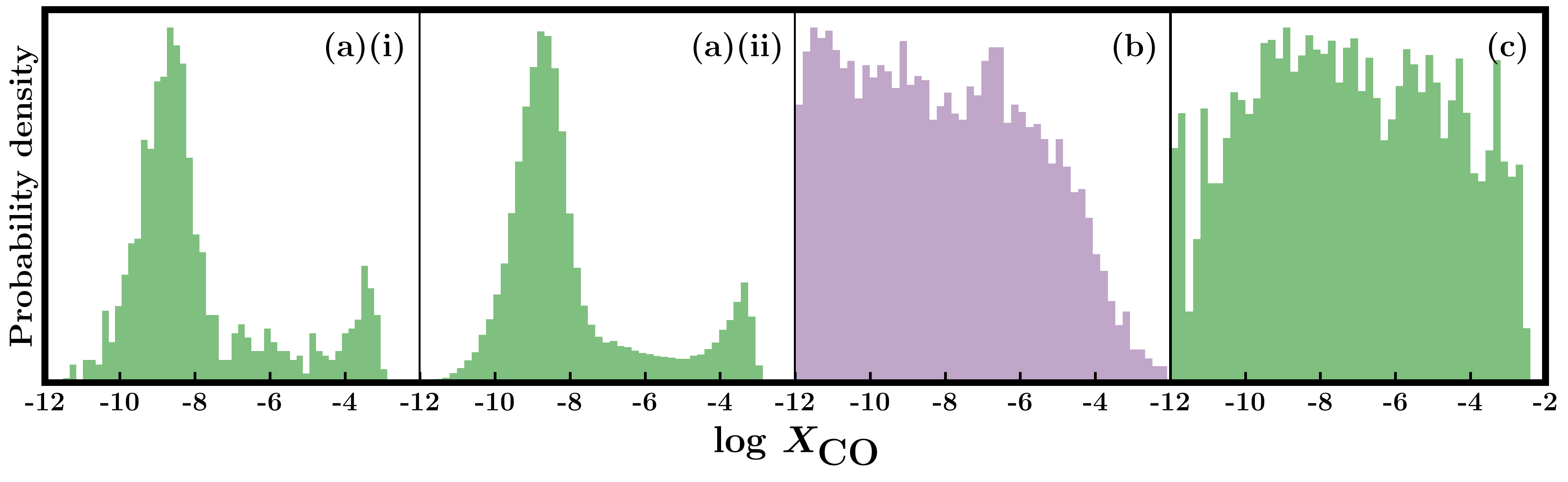}
    \caption{Marginal posterior distributions for CO abundance retrieved from the transmission spectrum of HD 209458b using different methods: (a)(i) Random Forest retrieval following the methods of \citetalias{PMN18}; (a)(ii) the same approach as (a)(i) but using 24000 trees instead of 1000; (b) Nested Sampling retrieval; (c) extended Random Forest retrieval. In order to aid comparison, the histograms have been scaled so that the bin with maximum probability in each plot has the same height. Given the lack of strong CO features in the spectral range of the data, a detection of CO is unexpected. Therefore the posterior in (a) is unphysical while those in (b) and (c) match with expectation.}
    \label{fig:co_post}
\end{figure*}

Next we take the same training set and apply our extended method as described in Section \ref{sub:extension}. More estimators are required to reach convergence in this case than when CO was not included; in this case convergence is reached after 24000 estimators had been trained. The results from this analysis are shown in Figure \ref{fig:rf_plus_hd_4param}. The marginalised posterior distributions of $T_{\textnormal{iso}}, P_{\textnormal{ref}}$ and $\log X_{\textnormal{H}_2\textnormal{O}}$ are once again very similar in shape to their Nested Sampling counterparts. The extended Random Forest method produces a broad distribution, leaving the CO abundance unconstrained as in the Nested Sampling case. Since the likelihood function for each estimator is evaluated directly, this method is able to infer that the value of CO abundance does not affect how well the model fits the data. This means that our algorithm does not suffer from the flaws described in \citet{Cobb_19} and is able to deal with unconstrained free parameters in such a way that a false constraint is avoided. Figure \ref{fig:co_post} shows a direct comparison of the marginal distributions for CO abundance obtained in each of the retrieval studies. We also show the CO posterior for a retrieval following the methods of \citetalias{PMN18}, but using the same number of trees as in our extended method. The spurious peak around $-8.5$ is still present, indicating that increasing the number of trees alone is not enough to solve this problem.

\section{Discussion and Conclusions} \label{section:discussion}

It has previously been suggested that using machine learning to perform retrievals could significantly reduce computation time, since a trained machine learning algorithm can make predictions extremely rapidly. The Generative Adversarial Network presented by \citet{zingales_2018} can predict model parameters from a spectrum in approximately 2 minutes, and \citet{Cobb_19} state that their approach can provide predictions in 1.5 seconds. The prediction time is not reported in \citetalias{PMN18}, but we find that 1000 estimators take a few seconds to make predictions for each retrieval considered in this study.

These numbers ostensibly suggest that machine learning retrievals are much faster than traditional methods, which can often take up to several hours depending on the size of the parameter space. However, these figures do not include the time taken to produce a training data set, nor do they incorporate the time taken to train the machine learning algorithm. According to \citet{zingales_2018} the training phase of their GAN using a forward model with 7 free parameters takes approximately three days per epoch on 20 CPU cores or about 9 hours per epoch on a GPU. The authors do not report how many epochs of training were required to fully train the network, nor do they say how long it took to generate the grid of $10^7$ models that were used for the training. \citet{Cobb_19} do not produce a unique data set for their retrievals, but instead use the same training set that was used in \citetalias{PMN18}. Each of their models takes approximately 20 minutes to train.

For the present study we compare the full retrieval duration of Nested Sampling and extended Random Forest retrievals for both the three- and four-parameter cases presented in Section \ref{section:HD}. We conduct retrievals on a synthetic data set binned to different resolutions from $R=10$ to $1000$, using the same computational resources for each (parallelisation over 4 CPU cores). For the Random Forest retrieval we include the time taken to produce the training data set, the training itself and the prediction step, however we note that in general only one training set would be needed to retrieve multiple observation instances of the same target.

\begin{figure}
\includegraphics[width=\columnwidth]{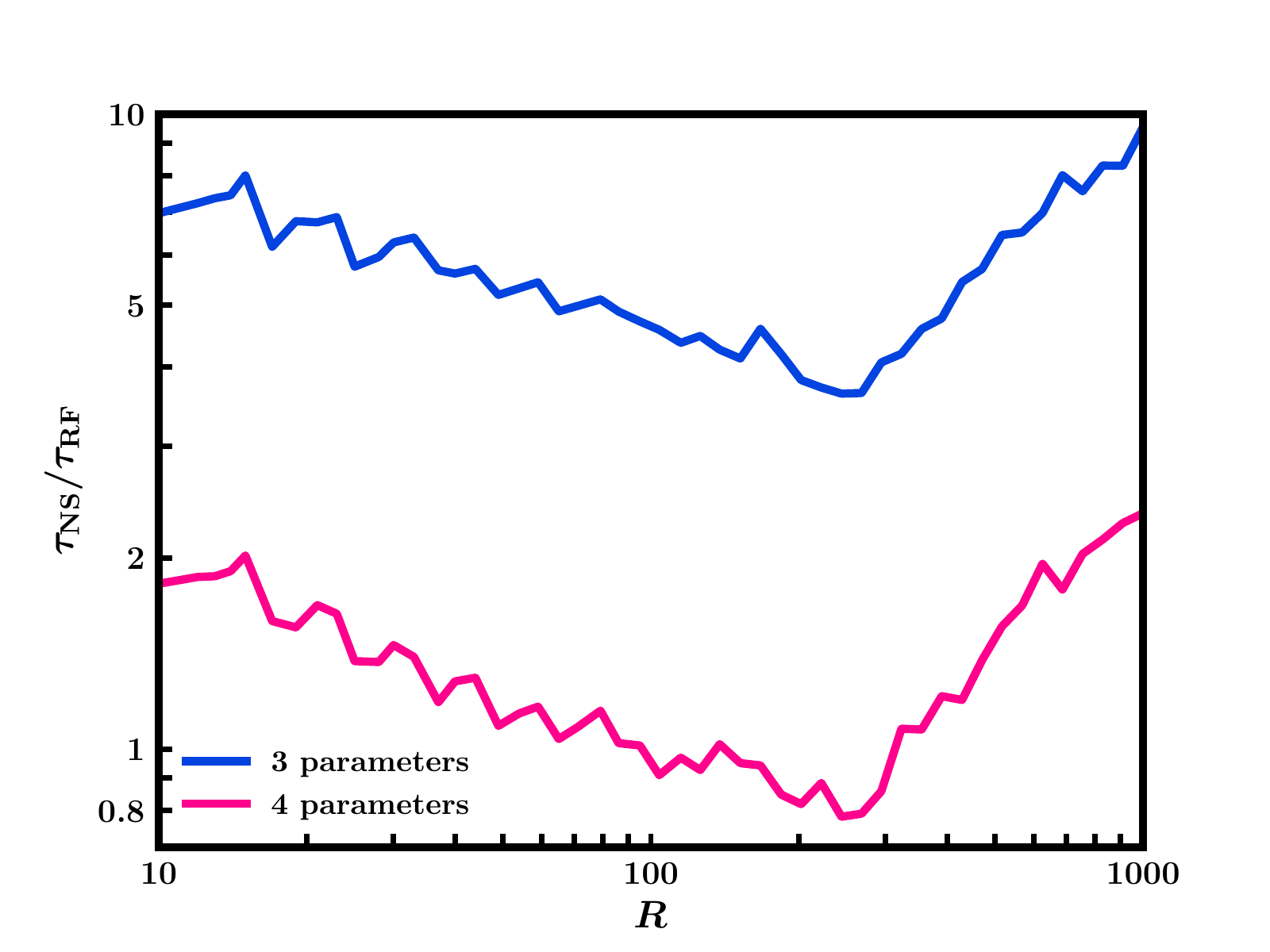}
    \caption{Comparison of the duration of Nested Sampling retrievals ($\tau_{\rm NS}$) versus extended Random Forest retrievals ($\tau_{\rm RF}$) as a function of spectral resolution $R$. Three-parameter Random Forest retrievals are faster than their Nested Sampling counterparts, whereas four-parameter Random Forest retrievals are only faster at low and high spectral resolutions, albeit only with a factor of two. For higher dimensions, Nested Sampling retrievals tend to be more efficient.}
    \label{fig:time_comparison}
\end{figure}

The results of this investigation are shown in figure \ref{fig:time_comparison}, where we show the relative speedup of the extended Random Forest retrieval compared to the Nested Sampling retrieval. In the three parameter case, the Random Forest retrieval always outperforms Nested Sampling by a factor of $\sim$4 to 8. At the resolution of the HD~209458b data used in Section \ref{sub:extension}, training time is approximately 4 seconds per 1000 estimators using the extended Random Forest approach. The biggest improvement over Nested Sampling is found at the lowest and highest wavelength resolutions, with a minimum at $R \sim 250$. While the duration of the Random Forest retrieval increases steadily with wavelength, following an approximate power law $\tau_{\rm RF} \sim R^{0.4}$, the duration of the Nested Sampling retrieval increases more slowly with $R$ up to $R \sim 250$ at which point $\tau_{\rm NS}$ increases rapidly. 

In the four parameter case, similar patterns are found in both types of retrieval, with both retrievals being slower overall. However, the addition of another parameter increases $\tau_{\rm RF}$ much more than $\tau_{\rm NS}$, resulting in retrievals of comparable duration. Training a Random Forest on high-dimensional data is much slower since a larger training set is required; in this case it takes about 80s to train 1000 estimators on 4 CPU cores. This indicates that increasing the number of free parameters and the size of the training set slows down the training significantly.

The applicability of a retrieval algorithm to higher-dimensional parameter spaces is an important factor to consider when comparing machine learning and traditional retrievals. As higher quality observed data and new line lists \citep[e.g.,][]{Exomol} become available it will be possible to search for an increasingly large number of atomic and molecular species, which will expand the number of possible free parameters in the forward model considerably. Additionally, extra parameters must be included to deal with other phenomena such as clouds \citep{Wakeford_15,Pinhas_clouds} and stellar heterogeneity \citep{pinhas_18}. We consider the feasibility of using the extended Random Forest method to perform a retrieval including additional chemical species and cloud/haze properties, following the prescription of \citet{MacDonald_17}. The model in this case requires 10 free parameters, and so we produce a large training data set consisting of $>10^6$ models. When a Random Forest is trained using this data, it predicts very few points with high likelihoods, suggesting that the parameter space is not sampled finely enough in the training set. This approach already requires far more model evaluations than a Nested Sampling retrieval using the same model, which converges after approximately 500000 model evaluations. A Random Forest retrieval with $n$ free parameters appears to require $\gtrsim 10^n$ models for an adequate training set. Full retrievals with optical and infrared data typically include up to $\sim$20  free parameters \citep{MacDonald_19,Madhu_20}, so a sufficient training set to carry out these retrievals using this method would be prohibitively large.  We therefore conclude that the Random Forest approach struggles to deal with higher dimensional parameter spaces efficiently. If a different algorithm can be used that performs more efficiently in high-dimensional parameter spaces and while still capturing uncertainties and degeneracies accurately, then it is possible that a machine learning approach could eventually compare to traditional methods for all cases, regardless of complexity.

At  present, each planet being analysed requires its own training data set for our extended Random Forest retrieval. An alternative approach to address the large computation time required for Random Forest retrievals in high-dimensional spaces might be to generate a large training set with many free parameters (including bulk parameters such as surface gravity) that would be applicable to a range of planets. This training set could then be used to train a model which could retrieve properties of spectra from different planets. Each retrieval would therefore only require the prediction step to be carried out after the training has been completed once. This kind of method has been applied to retrievals using deep learning: for example, \citet{Soboczenski_2018} used a large training set of $10^7$ models of terrestrial planet spectra which could be applied to numerous planets. \citet{zingales_2018} took a similar approach, using a training set of $10^7$ hot Jupiter spectra. In the next decade, with the advent of Ariel and the James Webb Space Telescope, we expect the number of planets with high-quality spectral data suitable for retrieval to increase significantly \citep{ARIEL}, so a machine learning approach that could apply to a range of planets may prove to be the most efficient option for conducting population studies of exoplanet atmospheres in the future.

Other than attempting to fully reproduce traditional retrievals, there may be other scenarios in which a machine learning-based approach could prove useful in this field: for example, a small number of predictions may be able inform the starting points for a MCMC retrieval, or could inform which molecules should be included in the full retrieval process, as discussed in \citet{waldmann_2016}. Additionally, while not explored in this study, the Random Forest algorithm provides information about the information content of each data point in the spectrum, and it was mentioned in \citetalias{PMN18} that this could be used to inform which wavelengths are most useful for future observations. We believe that combining machine learning algorithms with traditional methods can provide additional insight even when they are unable to replace existing methods entirely.

In this paper we have investigated the viability of using machine learning for atmospheric retrievals of exoplanets. We reproduced both the Nested Sampling and Random Forest results of \citetalias{PMN18} and we extended the methods from that paper so that the resulting posterior distribution from the Random Forest retrieval more closely matches that of the corresponding Nested Sampling retrieval. We applied this extended approach to a different planetary spectrum using a fully numerical forward model and found that once again we could accurately match the Nested Sampling and Random Forest retrievals. In addition, we found that our approach does not lead to spurious detections of parameters in cases where the parameter values should not be well-constrained, a problem found with the previous method. We have therefore developed a machine learning technique that can accurately and robustly reproduce the results of Bayesian retrievals. We investigated the potential for using this method to perform higher-dimensional retrievals and found that the algorithm requires a finely-sampled grid of training data in order to work well, making it prohibitively expensive to use this method in more complex cases. We conclude that while it is certainly possible to use machine learning techniques to reproduce traditional Bayesian retrieval results at least in low dimensions, the increased computational cost suggests that this approach does not yet provide a significant improvement on traditional methods. Future improvements in machine learning methodologies, as well as new strategies for applying these techniques to the present problem, will be required to surmount this challenge.

\section*{Acknowledgements}

We thank the anonymous referee for their helpful comments, which improved the quality of this manuscript. Matthew Nixon acknowledges support
from the Science and Technology Facilities Council (STFC), UK. This
research has made use of the NASA Astrophysics Data System, and the Python packages numpy, scipy, matplotlib and scikit-learn.




\bibliographystyle{mnras}
\bibliography{main} 







\bsp	
\label{lastpage}
\end{document}